\documentclass[manuscript,times]{aastex631}

\received{2022 December 19}
\accepted{2023 March 9}
\submitjournal{AJ}

\shorttitle{X-ray Emission from Nearby Low-Mass Exoplanet Host Stars}
\shortauthors{Brown et al.}

\begin{document}

\title{Coronal X-Ray Emission from Nearby, Low-Mass, Exoplanet Host Stars 
Observed by the MUSCLES and Mega-MUSCLES HST Treasury Survey Projects}

\correspondingauthor{Alexander Brown}
\email{Alexander.Brown@colorado.edu}

\author[0000-0003-2631-3905]{Alexander Brown}
\affiliation{Center for Astrophysics and Space Astronomy, 
University of Colorado, 389 UCB, 
Boulder, CO 80309, USA}

\author[0000-0002-5094-2245]{P. Christian Schneider}
\affiliation{Hamburger Sternwarte, 
Gojenbergsweg 112, 
D-21039, Hamburg, Germany}

\author[0000-0002-1002-3674]{Kevin France}
\affiliation{Laboratory for Atmospheric and Space Physics, 
University of Colorado, 600 UCB, 
Boulder, CO 80309, USA}

\author[0000-0001-8499-2892]{Cynthia S. Froning}
\affiliation{McDonald Observatory, 
University of Texas at Austin, 
Austin, TX 78712, USA}
\affiliation{Southwest Research Institute,  6220 Culebra Rd., 
San Antonio, TX 78238, USA}

\author[0000-0002-1176-3391]{Allison A. Youngblood}
\affiliation{Exoplanets and Stellar Astrophysics Lab, 
NASA Goddard Space Flight Center, 
Greenbelt, MD 20771, USA}

\author[0000-0001-9667-9449]{David  J. Wilson}
\affiliation{Laboratory for Atmospheric and Space Physics, 
University of Colorado, 600 UCB, 
Boulder, CO 80309, USA}

\author[0000-0001-5646-6668]{R. O. Parke Loyd}
\affiliation{School of Earth and Space Exploration, 
Arizona State University, Tempe, AZ 85287, USA}

\author[0000-0002-4489-0135]{J. Sebastian Pineda}
\affiliation{Laboratory for Atmospheric and Space Physics, 
University of Colorado, 600 UCB, 
Boulder, CO 80309, USA}


\author[0000-0002-7119-2543]{Girish M. Duvvuri}
\affiliation{Department of Astrophysical and Planetary Sciences, 
University of Colorado, Boulder, CO 80309, USA}
\affiliation{Center for Astrophysics and Space Astronomy, 
University of Colorado, 389 UCB, 
Boulder, CO 80309, USA}
\affiliation{Laboratory for Atmospheric and Space Physics, 
University of Colorado, 600 UCB, 
Boulder, CO 80309, USA}

\author[0000-0001-7458-1176]{Adam F. Kowalski}
\affiliation{National Solar Observatory, University of Colorado, 
3665 Discovery Drive, Boulder, CO 80303, USA}
\affiliation{Department of Astrophysical and Planetary Sciences, 
University of Colorado, Boulder, CO 80309, USA}
\affiliation{Laboratory for Atmospheric and Space Physics, 
University of Colorado, 600 UCB, 
Boulder, CO 80309, USA}

\author[0000-0002-3321-4924]{Zachory K. Berta-Thompson}
\affiliation{Department of Astrophysical and Planetary Sciences, 
University of Colorado, Boulder, CO 80309, USA}
\affiliation{Center for Astrophysics and Space Astronomy, 
University of Colorado, 389 UCB, 
Boulder, CO 80309, USA}

\begin{abstract}
The high energy X-ray and ultraviolet (UV) radiation fields of exoplanet  host stars play a crucial 
role in controlling the 
atmospheric conditions and the potential habitability of exoplanets. Major surveys of the 
X-ray/UV emissions from late-type (K and M spectral type) exoplanet hosts have been conducted by the 
MUSCLES and Mega-MUSCLES Hubble Space Telescope (HST) Treasury programs. These 
samples primarily consist of relatively old, ``inactive'',  low mass stars. In this paper we 
present results from X-ray observations of the coronal emission from these stars obtained using 
the Chandra X-ray Observatory, the XMM-Newton Observatory, and the Neil Gehrels Swift Observatory. 
The stars effectively sample the coronal activity of low-mass stars at a wide range of masses and ages. 
The vast majority (21 of 23) of the stars are detected and their X-ray luminosities measured. 
Short-term flaring variability is detected for most of the fully-convective (M $\leq$ 0.35 M$_{\odot}$) 
stars but not for the more massive M dwarfs during these observations. Despite this difference, 
the mean X-ray luminosities for these two sets of M dwarfs are similar with more massive 
(0.35 M$_{\odot}$ $\leq$ M $\leq$ 0.6 M$_{\odot}$) M dwarfs at 
$\sim$5 $\times$ 10$^{26}$ erg s$^{-1}$ compared to $\sim$2 $\times$ 10$^{26}$ erg s$^{-1}$ 
for fully-convective stars older than 1 Gyr. Younger, fully-convective M dwarfs have X-ray luminosities 
between 3 and 6 $\times$ 10$^{27}$ erg s$^{-1}$.
The coronal X-ray spectra have been characterized and 
provide important information that is vital for the modeling of the stellar EUV spectra.

\end{abstract}


\section{Introduction} \label{sec:intro}

Low mass K and M dwarfs are the focus of major research efforts to discover 
and characterize their exoplanet systems. It is easier to discover planets orbiting such low mass stars 
than for solar-like stars using both radial velocity and transit methods,
because the effects of the planets on the stellar signal is far larger.
Studies have shown the earth-like or super-earth planets are more common around M dwarfs and 
large gas-giant planets are rarer than for higher mass stars \citep{dressing15}. M dwarfs have very strong 
surface magnetic fields (see e.g. \citet{Shulyak_etal19}). 
Complex 3-6 kG magnetic fields fill and control the outer atmospheres of all M dwarfs 
\citep{Afram_Berdyuginal19}, leading to bright  X-ray emission 
from their coronae ( T $\geq$ 10$^6$ K) and strong ultraviolet emission lines from their 
chromospheres ( T $\sim$ 10$^4$ K) and transition region regions ( T $\sim$ 10$^5$ K), 
which can significantly influence conditions within planetary systems.  
 
Stellar radiation from each spectral region affects an exoplanet in different 
ways (\citet{ribas05}, \citet{france16} ). The optical/IR radiation dominates the radiated energy and 
controls the planet's lower atmospheric and surface heating. X-ray and EUV radiation 
at wavelengths shortward of 912 \AA\  play a major role in thermospheric heating and 
erosion of planetary atmospheres  \citep{gudel07}. FUV/NUV radiation controls the atmospheric chemistry via 
molecular formation and photolysis. The FUV radiation is dominated by emission in the  
1215.67 \AA\ H I Lyman-$\alpha$ line 
(\citet{france13}, \citet{Youngblood_etal16}).
The EUV region is a mixture of coronal and transition region emission lines that is 
not easily observable, because of both the obscuration imposed by the interstellar medium 
and a current lack of suitable observational capability in this spectral region, and thus 
must be reconstructed using spectral information from the X-ray and FUV regions \citep{duvvuri21}.

While considerable efforts have been devoted previously to studying young active M dwarfs 
in the X-ray and ultraviolet regions, until recently comparatively little was known about the 
activity levels of older ``inactive'' M dwarfs.  To remedy this lack 
the MUSCLES ['``Measurements of the Ultraviolet Spectral Characteristics of Low-mass 
Exoplanetary systems'': PI K. France: Program 13650: 125 HST orbits] \citep{france16} 
and Mega-MUSCLES [PI C. Froning: Program 15071: 157 HST orbits] 
(\citet{wilson21}, \citet{froning22})  HST Treasury programs\footnote{The high level science products 
of these programs, including panchromatiic SEDs, are archived at https://archive.stsci.edu/prepds/muscles/ }
have conducted an in-depth study of the UV and X-ray spectral energy distributions (SEDs) 
of K and M dwarf exoplanet hosts with a wide range of rotation periods and activity levels.
A sample totaling 23 stars with spectral types from K1 to M8 have been studied with HST UV 
observations and supporting X-ray observations from Chandra and XMM-Newton. 
The rotation rates of the stars range from still fast-rotating (few day period)
stars to older stars with $\sim$100 day rotation periods.  This is important because it 
is generally assumed that slower rotation links directly to lower magnetic activity levels.
Specific aims of these programs include characterizing the energetic radiation environment in 
the habitable zone of each star, measuring the flare properties on lower activity stars, and 
providing robust observational inputs to modeling the atmospheric photochemistry and 
the production of molecular tracers. 

The purpose of this paper is to provide a detailed overview of all the X-ray observations obtained 
by the MUSCLES and Mega-MUSCLES projects and to describe the coronal properties of the 
full sample. The methods used to characterize the X-ray data rely on standard X-ray 
modeling fitting techniques. Some earlier X-ray results from these projects have appeared in 
 \citet{Loyd_etal16},   \citet{Loyd_etal18}, and \citet{linsky20}.


\section{The MUSCLES and Mega-MUSCLES Sample of Low Mass Exoplanet Host Stars} \label{sec:sample}

 The MUSCLES and Mega-MUSCLES targets were chosen to provide spectral energy distributions 
 (SEDs) for stars that include multiple representative examples over spectral types from early K to late M 
 with a range of ages and activity levels. Many young K-M stars already existed  
 in the archives of UV and X-ray missions, but before these new studies observations of 
 older, low activity K-M dwarfs were notably lacking -- it was hard to convince proposal reviews 
 to observe ``boring'' stars. For maximum return the targets were selected from nearby 
 (almost all at distances less than 15 parsecs) stars that are mostly known exoplanet hosts, 
 and thus of individual significance for those studying and modeling exoplanetary systems.
 
 Our goal is to compare the measured coronal properties of our sample   
 with their fundamental physical properties such as mass, rotational period, and age. While 
 physical properties such as effective temperature, mass, and rotational period can be 
 measured fairly directly from observational data, others, particularly age, are much harder to 
 establish. Even directly determinable parameters can be confused in heterogeneous samples 
 where different methods are used to derive the values. Directly observable properties, 
 such as effective temperature as a proxy for mass and rotational period as a proxy for age,  
 can provide secure parameters for differential  comparison amongst our sample.
 The properties of the stars observed are outlined in Table \ref{table1}, including spectral types, 
 V magnitudes, distances, effective temperatures, masses, 
 rotation periods, and approximate ages.  We have aimed to use self-consistent methods for 
 deriving the physical parameters, and minimize extracting heterogeneous values from multiple 
 literature sources. Spectral types were obtained from a critical examination of published 
 values and are primarily from the PMSU survey (\citet{reid95}, \citet{hawley96}) and the 
 CARMENES \citep{alonso-floriano15} and MEarth \citep{newton14} target sample papers. 
The very well determined stellar distances are from the parallax data in the GAIA EDR3 catalog 
\citep{GAIA16, GAIA20}. Minimizing the uncertainties on these parameters allows 
 the stars to be placed in ordered sequences (e.g. into a reliable mass sequence) with 
 little ambiguity. 


\begin{deluxetable*}{cccccccccc}
\tablecaption{MUSCLES - Mega-MUSCLES Exoplanet Host Star Sample\label{table1}}
\tablewidth{0pt}
\tabletypesize{\footnotesize}
\tablehead{
\colhead{Star}  &\colhead{Spectral} & \colhead{V} &
 \colhead{Distance}  &\colhead{T$_{eff}$} &\colhead{T$_{eff}$} & \colhead{Mass} &
 \colhead{P$_{rot}$} & \colhead{Age}  &\colhead{Refs.} \\
 \colhead{ }  &\colhead{Type} &\colhead{ } &\colhead{} & \colhead{[Houdebine]}& \colhead{[Pineda]} &
 \colhead{  } &  \colhead{ }  & \colhead{[Approx.]}  &\colhead{ }   \\
 \colhead{ }  &\colhead{ } &\colhead{(mag) } &\colhead{(pc)} & \colhead{(K)}& \colhead{(K)} &
 \colhead{(M$_\odot$)}&  \colhead{(days)}  & \colhead{ (Gyr) }  &\colhead{ }   
}
\decimalcolnumbers
\startdata
HD 97658 &K1 V  & 7.7&21.563$\pm$0.010  & 5155 & \nodata &0.77$\pm$0.05 & 34$\pm$2    &  3.8$\pm$2.6 & 1, 2  \\
$\epsilon$ Eri & K2 V& 3.7&3.220$\pm$0.001& 5041 & \nodata &0.83$\pm$0.01&  11.68            & 0.4$^{+0.4}_{-0.2}$  & 3, 4\\
HD 40307 &K2.5 V& 7.1&12.932$\pm$0.003 & 4963 &  \nodata &0.77$\pm$0.05 &31.8$\pm$6.7 & 7.0$\pm$4.2 & 5, 2\\
HD 85512 &K6 V  & 7.7&11.277$\pm$0.002  & 4451 &  \nodata &0.62$\pm$0.01 &45.9$\pm$0.4  & 8.2$\pm$3.3& 5, 2\\
GJ676A    &M0 V  & 9.6 &15.980$\pm$0.008 & 3838 & 4014$^{+94}_{-90}$      &0.631$\pm$0.017& 41.2$\pm$3.8  &  \nodata  &  5    \\
GJ 649    &M1 V  & 9.7  &10.391$\pm$0.003 & 3705 & 3621$^{+41}_{-40}$      & 0.524$\pm$0.012&  23.8$\pm$0.1  & 4.5$^{+3.0}_{-2.0}$ & 6, 7  \\
GJ 832     &M1.5 V  & 8.7 &4.967$\pm$0.001& 3590 & 3539$^{+79}_{-74}$       &0.441$\pm$0.011& 45.7$\pm$9.3  &  (8.4 )    & 5, 8   \\
GJ 667C  &M1.5 V&10.2 &7.243$\pm$0.002 & 3570 & 3443$^{+75}_{-71}$       &0.327$\pm$0.008& 103.9$\pm$0.7 &$\ge$ 5  &5, 9   \\
GJ 15A    &M2 V  & 8.1  &3.562$\pm$0.001   & 3656 & 3601$^{+12}_{-11}$      & 0.393$\pm$0.009&  30.5                 &   $\sim$3       & 10, 15 \\
GJ 176    &M2 V &10.0&9.485$\pm$0.002   & 3542 & 3632$^{+58}_{-56}$       &0.485$\pm$0.012& 38.9                 &  8.8$^{+2.5}_{-2.8}$    & 11, 7  \\
GJ 436    &M3 V  & 10.6 &9.775$\pm$0.003  & 3464 & 3477$^{+46}_{-44}$       &0.425$\pm$0.009& 44.6$\pm$0.2  &  8.9$^{+2.3}_{-2.1}$    & 6, 7  \\
GJ 674    &M3 V  &   9.4 &4.553$\pm$0.001  & 3478 & 3404$^{+59}_{-57}$       &0.353$\pm$0.008& 33.3              &    0.20   &   11, 12 \\
GJ 581    &M3 V  & 10.6 &6.301$\pm$0.001  & 3423 & 3424$^{+43}_{-42}$       &0.307$\pm$0.007&132.5$\pm$6.3 &  6.6$^{+2.9}_{-2.5}$    & 5, 7 \\
GJ 163    &M3.5 V& 11.8 &15.135$\pm$0.004& 3413 & 3460$^{+76}_{-74}$      &0.405$\pm$0.010& 61.0$\pm$0.3  &  1-10    & 5, 13 \\
GJ 849   &M3.5 V & 10.4 &8.815$\pm$0.002  & 3408 & 3492$^{+70}_{-68}$      &0.465$\pm$0.011&  32.2$\pm$6.3  & 4.9$^{+3.0}_{-2.1}$    &  5, 7    \\
GJ 3843 &M3.5 V & 13.0 &13.373$\pm$0.006 & 3353 & 3278$^{+74}_{-70}$      &0.232$\pm$0.006 & 91.4           &  \nodata   &  16    \\
GJ 876A &M4 V& 10.2 &4.672$\pm$0.001   & 3304 & 3201$^{+20}_{-19}$      &0.346$\pm$0.007& 81.0$\pm$0.8   & 8.4$^{+2.2}_{-2.0}$    & 6, 7 \\
GJ 729    &M4 V& 10.4 &2.976$\pm$0.001  & 3276 & 3248$^{+68}_{-66}$      &0.177$\pm$0.004 & 2.848$\pm$0.001&  0.7$^{+0.5}_{-0.3}$  &  14, 10, 15\\
GJ 1132  &M4 V & 13.5 &12.607$\pm$0.003  & (3270)& 3196$^{+71}_{-70}$     &0.194$\pm$0.005 & 129            & $\ge$ 5   & 16,  17 \\
GJ 699   &M4 V    & 9.5 &1.828$\pm$0.001    & 3266 & 3223$\pm$17     &0.161$\pm$0.004 & 145$\pm$15      &   7-10   & 18, 19  \\
LHS 2686&M4.5 V  & 14.6 &12.189$\pm$0.005&(3157)& 3119$^{+70}_{-68}$    &0.157$\pm$0.004 & 28.8       &   \nodata    & 10  \\
GJ 1214  &M4.5 V  & 14.7 &14.642$\pm$0.014  & 3150 &  3111$^{+69}_{-66}$    & 0.181$\pm$0.005 &125$\pm$5  &    6-10  & 20 \\
TRAPPIST-1&M8 V& 18.8 &12.467$\pm$0.011& (2628)& 2619$^{+71}_{-66}$  &0.090$\pm$0.003& 3.295$\pm$0.003&7.6$\pm$2.2&  21, 22  \\
\enddata
\tablecomments{Rotational period and Age references:  1  \citet{guo20}, 2 \citet{bonfanti15}, 3 \citet{donahue96}, 
4 \citet{janson08}, 5 \citet{mascareno15}, 6 \citet{diez_alonso19}, 7 \citet{veyette18}, 8 \citet{bryden09}, 9 \citet{anglada-escude13},  
10 \citet{newton16}, 11 \citet{kiraga07}, 12 \citet{montes01},  13  \citet{bonfils13}, 14 \citet{Ibanez_Bustos20}, 
15 \citet{allen98}, 16  \citet{newton18}, 17 \citet{berta-thompson15}, 18 \citet{toledo-padron19}, 19 \citet{ribas18}, 20 \citet{mallonn18}
21 \citet{Vida17}, 22 \citet{burgasser17}
 }
\end{deluxetable*}

\subsection{Effective Temperatures}
The stellar effective temperatures range from early K stars at $\sim$5000 K to mid-M stars 
at  $\sim$3000 K. The M8 V star TRAPPIST-1 is the coolest star in the sample at just above 2600 K.
The effective temperatures listed are taken primarily from \citet{houdebine19}, which provides 
a uniform temperature scale based on the Cousins (R-I)$_C$ color index for most (17 of 23) 
of the stars in our sample. \citet{houdebine19} provide a detailed comparison to  
earlier temperature determinations in the literature. Systematic differences are common between 
different method of temperature determination and can easily be  $\sim$100-200K. 
The 1$\sigma$ uncertainty on individual \citet{houdebine19} temperatures is $\sim$40-50 K.
The effective temperatures for three 
additional stars (HD 97658, HD40307, and GJ676A) were estimated from the \citet{houdebine19} 
calibration using (R-I)$_C$ photometry from \citet{Koen_etal10}. The temperatures derived for 
HD 97658 and HD40307 were 5155 K and 4963 K respectively, which are comparable to literature 
values of 5162$\pm$6 K \citep{kulen17}, 5175 K \citep{takeda20}  and 5192  K \citep{guo20} 
for HD 97658 and 4956 K \citep{tuomi13} for HD 40307.  \citet{Schweitzer_etal19} used spectral fitting 
of high resolution optical spectra to derive the physical parameters of their CARMENES sample 
of M dwarf exoplanet hosts and list effective temperatures for eight stars in our sample that 
are also included in \citet{houdebine19}. The mean difference between the Houdebine et al.  and 
Schweitzer at al. values is -25 K, with a standard deviation of 35 K, and shows that there is general 
consistency between different methods of temperature estimation.
Effective temperatures for the three remaining stars in our X-ray sample were taken from the following sources: 
GJ1132 \citep{berta-thompson15}, LHS2686 \citep{Schweitzer_etal19}, 
and TRAPPIST-1  \citep{Gonzales_etal19}.  

For comparison, effective temperatures estimated by \citet{pineda21b} are also presented. Rather than being 
derived directly from an observable quantity, these values are the end product of a modeling sequence 
consisting of mass derived from K-band absolute magnitude, radius from a mass-radius relation, and 
bolometric luminosity from the combination of J band magnitude and bolometric correction plus distance 
from parallax. This modeling framework is applicable in the mass range 0.09 - 0.7 M$_{\odot}$ and are thus 
applicable to the M dwarfs in our sample but not the early K dwarfs. 
These effective temperatures form a self-consistent sequence with the mass estimates 
discussed below.

\subsection{Masses}
The observed stars sample 
the mass range 0.09 - 0.83 M$_{\odot}$, with the greatest concentration covering the early- and mid-M 
stars between 0.1 and 0.5 M$_{\odot}$.
M dwarf masses can also be estimated directly from an observable quantity using the correlation between 
K band absolute magnitude and mass, with metallicity playing only a minor role (see e.g. \citet{mann19}).
In Table \ref{table1} the mass estimates derived by \citet{pineda21b}, based on the calibration of 
\citet{mann19}, are presented, but masses for the 4 K dwarfs are taken from \citet{vanGrootel14} [HD 97658], 
\citet{sousa08} [HD 40307], and \citet{bonfanti15} [$\epsilon$ Eri, HD 85512]. The division between stars 
with radiative cores and fully convective stars occurs at a mass of 0.35 M$_{\odot}$ \citep{chabrier97}.

\subsection{Rotational Periods}
A wide range of rotational periods are covered by the sample stars -- a few fast rotators with periods of 
a few days, many stars with intermediate periods of 20-60 days, plus several very slowly rotating stars 
with periods as long as $\sim$150 days. This range in rotation should ensure that all levels of coronal activity 
are sampled.
Stellar rotational periods can be derived from long (multi-year) photometric and spectroscopic time series, 
even for slowly rotating, inactive M dwarfs. Generally, photometric data provides clearer signals with 
long-lived dark starspot groups modulating the light curve. Rotational signatures in 
chromospheric emission line spectra, such as Ca II H\&K and H Balmer $\alpha$, are more difficult to establish. 
For this reason, we give preference to rotational periods derived from photometry in Table \ref{table1}. 
These rotation periods relate to the rotation at the latitudes where the strong magnetic field regions emerge 
through the stellar photosphere, which is potentially far from the stellar equator. Therefore, it is important 
not to assume that the measured periods represent the equatorial rotation, given that much remains to be 
learned about differential rotation on low mass stars.

\subsection{Ages}
Determining stellar ages is a nontrivial problem, particularly for M dwarfs. Commonly, when considering 
magnetic activity among stellar 
age samples, many papers settle for merely dividing the stars into ``young'' and ``old'' stars with the 
divide at $\sim$1 Gyr. For our study one goal is to understand how stellar rotation and 
the consequent magnetic activity change with age, hence it is important to have independently determined ages 
that are not estimated based on stellar activity indicators (such as Ca II HK or X-ray emission) or on rotational 
periods themselves. The ages listed are from a heterogeneous set of methods, including Galactic kinematic 
and elemental evolution studies. Evolutionary tracks provide little scope for age separation except for the very 
youngest stars.  Three stars in the MUSCLES/Mega-MUSCLES sample are definitely younger than 1 Gyr 
($\epsilon$ Eri, GJ 674, GJ 729). Most of the other stars are solar age or older. 

\subsection{GJ 15B}
GJ 15B (M3.5 V) forms a wide binary with GJ 15A and is also detected in the 
GJ 15A Chandra observation. The GJ 15B X-ray source is located 
34 arcseconds NE of A. Although GJ 15B is not a known exoplanet host and, thus, is not 
included in the MUSCLES/Mega-MUSCLES sample, for completeness 
the properties of its X-ray emission are 
included in the results presented later in this paper. It is a slow rotator with a period of 
167.6 days \citep{newton16}. \citet{pinamonti18} discuss the detailed properties of 
the GJ 15 binary system and provide effective temperature and mass estimates for 
GJ 15B of 3304$\pm$70 K and 0.15$\pm$0.02 M$_\odot$ . The temperature estimate 
is typical for an M3.5 dwarf but the mass is perhaps a little low based on 
other stars with this spectral type. Their estimate 
 for the bolometric luminosity is 3.3 $\times$ 10$^{30}$ erg s$^{-1}$, but 
 with a very large uncertainty of a factor of 16.6. Using the observed K band magnitudes 
 for the two stars (A: K=4.02; B: K=5.95) implies a luminosity ratio of 5.9 and, 
 based on the well determined bolometric luminosity for GJ 15A, leads to an 
 improved bolometric luminosity of 1.44 $\times$ 10$^{31}$ erg s$^{-1}$ for B, 
 which falls within the error range of \citet{pinamonti18}. This larger bolometric luminosity 
 and a mass of 0.15 M$_\odot$  are very similar to those of the M4 V stars 
 in Table \ref{table1} and imply that GJ 15B is a fully convective M4 star. 

\vspace{0.5cm}

\section{X-ray Observations} \label{sec:xray-obs}

Characterization and modeling of the high energy radiation field surrounding exoplanet host stars cannot 
be successful using UV spectra alone; therefore, when constructing our HST Treasury program surveys 
it was clear that dedicated X-ray observations were essential to achieve the science goals of the larger 
projects. A mixture of new dedicated and archival observations were obtained using the Chandra,
XMM-Newton, and Swift X-ray observatories. 
Details of the X-ray observations of these targets are outlined in Table \ref{table2}.  

\begin{deluxetable*}{cccccc}
\tablecaption{MUSCLES - Mega-MUSCLES X-ray Observations\label{table2}}
\tablewidth{0pt}
\tabletypesize{\footnotesize}
\tablehead{
\colhead{Star}  &{Spectral} & \colhead{Instrument}& \colhead{Exposure} & 
\colhead{Observation Date - } & \colhead{ObsID} \\
 &\colhead{Type} & \colhead{ }&\colhead{Time (ks)} & \colhead{Start Time [UT]} &  }
\startdata
HD 97658 &K1 V  & ACIS & 12.9  &2015-10-17 13:37:51& 16668 \\ 
\nodata& \nodata   & ACIS & 19.2  &2015-12-11 12:39:18 & 18724 \\
\nodata& \nodata   & ACIS & 19.2  &2016-03-05 15:59:55 & 18725 \\
$\epsilon$ Eri &K2 V& EPIC&4.9(pn);10.5(MOS)& 2015-02-02 08:16:05 & 0748010101\\
HD 40307 &K2.5 V& EPIC&12.5(pn);5.0(MOS)&2015-03-17 02:47:53 & 0763810101\\
HD 85512 &K6 V  & EPIC&14.57& 2015-05-24 10:09:51 & 0763810201\\
GJ676A  & M0 V  & Swift  & 6.71  & 2018-06-19/2019-04-1/2  & See Sect. \ref{subsec:swift-obs} \\
GJ 649   &M1 V   & EPIC& 18.0  &2018-03-03 13:44:39 & 0810210401 \\
GJ 832   &M1.5 V& EPIC  &8.9(pn);11.5(MOS) & 2015-10-11 02:50:32 & 0748010201   \\
GJ 667C &M1.5 V&ACIS & 9.1/18.2  &2015-08-07 04:43:16/ 14:10:52& 17318/17317 \\
GJ 15A   &M2 V& ACIS & 23.8  &2019-02-12 17:25:17 & 20617 \\
GJ 176   &M2 V &ACIS  & 9.6  &2015-02-28 00:13:08 & 17320 \\
\nodata&\nodata& ACIS & 19.6 &2015-03-02 03:03:26 & 17319 \\
GJ 436   &M3 V&ACIS& 18.8 &2013-02-16 14:51:23 & 14459 \\
\nodata&\nodata& ACIS & 18.9 &2013-04-18 08:31:13 & 15537 \\
\nodata&\nodata& ACIS & 19.8 &2013-06-18 06:55:32 & 15536 \\
\nodata&\nodata& ACIS & 18.8 &2014-06-23 10:42:32 & 15642 \\
\nodata&\nodata& ACIS &   9.1 &2015-06-25 02:26:52 & 17322 \\
\nodata&\nodata& ACIS & 19.6 &2015-06-26 01:22:02 & 17321 \\
GJ 674   &M3 V  & EPIC &30.4 &2018-04-03 06:14:25 & 0810210301   \\
GJ 581   &M3 V  & ACIS& 47.6 &2014-06-12 08:19:02 & 15724   \\
GJ 163   &M3.5 V & ACIS &28.6 &2019-03-04 18:24:00 & 20621   \\
GJ 849   &M3.5 V  & ACIS & 28.0 &2019-06-14 23:15:36 & 20620   \\
GJ 3843   &M3.5 V  & ACIS  &6.9 &2016-12-16 19:10:13 & 18938  \\
GJ 876A   &M4 V  & ACIS & 9.9  &2015-06-04 17:18:25 & 17316 \\
\nodata&\nodata& ACIS & 19.8 &2015-06-05 05:15:14 & 17315 \\
GJ 729   &M4 V  & EPIC &23.0 &2018-04-20 04:59:03 & 0810210201   \\
GJ 1132   &M4 V  & EPIC &46.6 &2019-01-10 10:06:49 & 0804930201  \\
GJ 699   &M4 V  & ACIS & 26.7  &2019-06-17 08:51:05 & 20619 \\
LHS 2686  &M4.5 V  &ACIS & 26.7  &2019-03-08 11:56:58 & 20618 \\
GJ 1214    &M4.5 V  & ACIS & 30.5 &2014-06-17 03:20:29 & 15725 \\
TRAPPIST-1&M8 V& EPIC& 24.6  &2018-12-10 03:52:57 & 0810210101\\
\enddata
\end{deluxetable*}

\subsection{Chandra Observations} \label{subsec:chandra-obs}
Thirteen stars were observed by Chandra using the ACIS-S3 detector.
Dedicated observations were obtained including 
two (GJ 581, GJ 1214) by proposal 15200539 (PI Brown), four (GJ667C, 
GJ 176, GJ 436, GJ 876A) by proposal 16200943 (PI France; MUSCLES), 
five (GJ 15A, GJ 163, GJ 849, GJ 699, LHS 2686) by proposal 19200772 
(PI Froning; Mega-MUSCLES). Archival observations were used for 
HD 97658 (programs 16200348 PI. Miller and 16208530 PI Wheatley),
GJ 463 (program 14200978, PI Ehrenreich), and GJ3843 (=L980-5, 
program 81200661, PI Wright).

\subsection{XMM-Newton Observations} \label{subsec:xmm-obs}
Nine stars were observed by XMM using the EPIC detectors. 
Dedicated observations were obtained including 
two (HD 40307, HD 85512) by proposal 076381 (PI Brown), 
two ($\epsilon$ Eri, GJ 832) by proposal 074801 (PI France; MUSCLES), 
four (GJ 649, GJ 674, GJ 729, TRAPPIST-1) by proposal 081021 
(PI Froning; Mega-MUSCLES). Additional archival observations were used for 
GJ 1132 (program 080493, PI King) and TRAPPIST-1 (program 080444, PI Wheatley).

\subsection{SWIFT Observations} \label{subsec:swift-obs}
No Chandra or XMM-Newton observations exist for Mega-MUSCLES 
target GJ 676A, and we have used archival Swift XRT 
observations to measure the X-ray 
activity level for this star. The two most useful data sets were obtained 
on 2018 June 19 (ObsIDs: 0001078003-4; 3157 s exposure) and 
on 2019 Apr 1/2 (ObsIDs: 0001078006-9; 3554 s exposure). 
Combined these observations provide a background-corrected total of 
32.6 source counts and a source count rate of 4.86$\pm$0.85 10$^{-3}$ 
ct s$^{-1}$ (5.7 $\sigma$ detection). The individual source count rates for 
the 2018 and 2019 data were 5.68$\pm$1.34 10$^{-3}$ 
ct s$^{-1}$ (4.2 $\sigma$ detection) and 4.1$\pm$1.1 10$^{-3}$ 
ct s$^{-1}$ (3.8 $\sigma$ detection) respectively. There is no evidence for 
significant source variability, although  we note that the number of counts is very 
limited.

\section{X-ray Data Analysis}\label{sec:xray-analysis}

The X-ray data were analyzed with standard analysis techniques for 
the various X-ray detectors used. Timing and spectral studies were conducted when the 
 number of detected source counts was sufficient. Analysis was typically 
performed on standard pipeline-processed datasets obtained 
from the mission archives.
The X-ray spectra were fitted using the XSPEC software package
(Versions 12.5.1 - 12.12.0 -- \citet{arnaud96, dorman03}). The spectral 
parameterization provides estimates of the characteristic coronal temperature 
and the associated volume emission measure. Depending on the number of 
source counts collected either one-temperature (1T), two-temperature (2T), 
or in some cases even three-temperature (3T) parameterizations using VAPEC 
spectral models  \citep{smith01} were possible.
All the stars observed are very close (mostly less than 15 pc) to the Sun
and thus have very low intervening hydrogen column densities (N$_H$), which
cannot be constrained by CCD-resolution {\it Chandra} and
{\it XMM-Newton} spectra. Therefore, we adopted a fixed value of
N$_H$, typically 1 $\times$ 10$^{19}$ cm$^{-2}$, when fitting the spectra, rather than 
attempting to derive N$_H$ directly from the X-ray data.

The results from analysis of our X-ray observations are
presented in Tables \ref{table3} and  \ref{table4} for the XMM-Newton observations 
and Tables  \ref{table5} and  \ref{table6} for the Chandra observations. 
These tables provide information on the detected source counts and the
background-corrected count rate,  presence or absence of
significant source variability,  the observed 0.3-10 keV flux, the X-ray luminosity
and X-ray to bolometric luminosity ratio in the same energy range,
and the derived coronal temperature (kT$_1$) and volume
emission measure (VEM$_1$)  for the dominant thermal component, 
the temperature (kT$_2$) and volume emission measure (VEM$_2$)
for any identifiable secondary thermal component, and the
reduced $\chi^{2}$ for the fitting parameterization. 
For variability, a question mark indicates 
uncertainty, while parentheses indicate long-term variability.

\subsection{Chandra Observations}\label{subsec:chandraobs}

The ACIS data were analyzed using standard CIAO \citep{CIAO06} software tools  
on the standard pipeline-processed dataset obtained from the mission archive.
Source events from the ACIS observations were extracted using
extraction circles with radii of 5 CCD pixels (2.5 arcseconds). 

A barycentric correction was applied to the times in each event list using the  
CIAO tool {\it axbary}. Each dataset was tested for source variability 
using the CIAO tool {\it glvary}. This 
tool searches for variability using the Gregory-Loredo algorithm, which tests for
nonrandom bunching of the event times across multiple time bins. {\it glvary} is the 
standard tool used to test for variability in the major Chandra source catalogs.
{\it glvary} produces a variability index (VARINDEX) which ranges from 0 to 10, with 
values of 5 or above indicating a variable source.

\subsection{XMM-Newton Observations}\label{subsec:xmmobs}
XMM-Newton observations were obtained using primarily the EPIC pn and MOS detectors. 
Background particle radiation can seriously degrade the EPIC data and high background 
time intervals were removed from the data before spectral analysis but not for light curve 
construction. Source extraction circle sizes are usually around 15 arcsec, sometimes slightly 
adjusted for very weak or strong sources. The background was estimated from detector 
regions on the same detector chip, which are much larger than the source size. The spectra 
from the pn and the two MOS detectors were jointly analyzed with XSPEC. 
Because flare signatures are usually well visible in the light curves, no formal variability analysis 
has been performed.

\subsection{Swift Observations}\label{subsec:swiftobs}
Too few counts were detected for GJ 676A by Swift for any detailed 
spectral fitting. Therefore, we assume a coronal temperature of 0.24 keV 
(2.8 MK), typical for most of the other quiescent early-mid M dwarfs 
in our sample,  sub-solar abundances (0.4 solar) and a 
interstellar hydrogen column of 10$^{19}$ cm$^{-2}$ as inputs for 
estimating its X-ray flux and luminosity. The X-ray flux was estimated using 
WebPIMMS\footnote{The Portable Interactive Multi-Mission Simulator,  
https://heasarc.gsfc.nasa.gov/cgi-bin/Tools/w3pimms/w3pimms.pl} and 
APEC spectral models \citep{smith01} .
The assumed abundance has only a small influence on the estimated flux with 
abundances of 0.2 solar and 1.0 solar producing a 4\% flux decrease and a 4\% flux increase 
respectively. The assumed temperature has a larger effect. Increasing the temperature 
has less effect than lowering it. Doubling the temperature to 0.48 keV (5.6 MK) increases 
the flux by only 5\%, but decreasing the temperature to 0.15 keV (1.8 MK) leads to a flux 
estimate 19\% smaller.

\begin{deluxetable*}{cccccc}
\tabletypesize{\footnotesize}
\tablecaption{MUSCLES K1-M0 Dwarf Datasets and Measured Coronal X-ray Properties \label{table3}}
\tablewidth{0pt}
\tablehead{
\colhead{ }  & \colhead{$\epsilon$ Eri} & \colhead{HD 40307} &
 \colhead{HD 85512} &  \colhead{HD 97658} & \colhead{GJ 676A} }
\startdata
Instrument&EPIC\tablenotemark{a}  & EPIC\tablenotemark{a}  & EPIC\tablenotemark{a}   & ACIS  & SWIFT-XRT\\
Source Counts (ct) & $3.8\times10^4$ & 71 & 283    & 43   & 32 \\
Count Rate (ct s$^{-1}$) &$6.05\pm0.03$  & $(4.2\pm0.7)\times10^{-3}$ & $(16.6\pm1.4)\times10^{-3}$  &
 $(8.3\pm1.4)\times10^{-4}$  & $(4.9\pm0.9)\times10^{-3}$ \\ 
Variable (Yes/No) & (Y)\tablenotemark{b}  & N & N & N?  &  N \\
f$_{X}$ (erg cm$^{-2}$ s$^{-1}$)  & (9.2$\pm$0.1)$\times$10$^{-12}$ & (7$\pm$2)$\times$10$^{-15}$ & (1.9$_{-0.3}^{+0.4}) \times$10$^{-14}$ 
& (7.7$\pm$1.3)$\times$10$^{-15}$ &  (9.7$\pm$1.7)$\times$10$^{-14}$ \\
$\log $L$_{X}$  (erg s$^{-1}$)  & $28.06\pm0.06$ & $26.15\pm0.15$ & $26.46\pm0.09$ &$26.63\pm0.08$  &  $27.47\pm0.08$  \\
log L$_{X}$/L$_{bol}$   &   $-5.03\pm0.06$ &   $-6.80\pm0.13$   & $-6.26\pm0.09$  &  $-6.59\pm0.08$ & $-5.06\pm0.08$  \\
N$_{H}$ (10$^{20}$ cm$^{-2}$)  & $0.5_{-0.5}^{+1.1}$ & 0.1\tablenotemark{c} & 0.1\tablenotemark{c} &0.1\tablenotemark{c} & 0.1\tablenotemark{c}\\
kT$_{1}$ (keV)   & $0.12_{-0.02}^{+0.09}$ & $0.15\pm0.06$ & $0.25_{-0.03}^{+0.04}$ & $0.23\pm0.04$ & \nodata  \\
VEM$_{1}$ (10$^{49}$ cm$^{3}$) & $53_{-19}^{+45}$ & $1.2_{-0.6}^{+3.0}$
& $2.3\pm0.3$ &  $2.0\pm1.3$ & \nodata   \\
kT$_{2}$ (keV)      & $0.32\pm0.01$ & \nodata & \nodata   & \nodata  & \nodata  \\
VEM$_{2}$ (10$^{49}$ cm$^{3}$) & $78.4_{-10}^{+11}$ & \nodata & \nodata  & \nodata  & \nodata   \\
kT$_{3}$ (keV)        & $0.70_{-0.07}^{+0.04}$ & \nodata & \nodata  & \nodata  & \nodata  \\
VEM$_{3}$ (10$^{49}$ cm$^{3}$) & $22.2_{-3.5}^{+5.1}$ & \nodata & \nodata  & \nodata  & \nodata  \\
Red. $\chi^{2}$ & 1.28\tablenotemark{d} & 1.00 & 0.95\tablenotemark{e} &  1.95 & \nodata  \\
\enddata
\tablenotetext{a}{For XMM-Newton summed EPIC (pn+mos1+mos2) -- Fit performed using C-statistic, $\chi^2$ provided for reference only.}
\tablenotetext{b}{No significant variability during the MUSCLES EPIC-pn observation but long-term variability present (see Sect. \ref{subsec:Kstars}).}
\tablenotetext{c}{Fixed $N_H=1\times10^{19}\,$cm$^{-2}$ for the fit.}
\tablenotetext{d}{Adjusted abundances: Low-FIP (Fe, Mg): $0.47_{-0.06}^{+0.08}$, Mid-FIP (C, N, O, Si, S): $0.39_{-0.11}^{+0.09}$, High-FIP (Ne, Ar): $0.46_{-0.13}^{+0.09}$}
\tablenotetext{e}{Iron abundance fixed to 0.05; upper limit 0.16.}
\end{deluxetable*}

\begin{deluxetable*}{ccccccc}
\tabletypesize{\footnotesize}
\tablecaption{XMM-Newton M Dwarf Datasets and Measured Coronal X-ray Properties \tablenotemark{a} \tablenotemark{b} \tablenotemark{c}  \label{table4}}
\tablewidth{900pt}
\tablehead{
\colhead{ }  &  \colhead{GJ 832} &\colhead{GJ 649} & \colhead{GJ 674} &
\colhead{GJ 729} &    \colhead{GJ 1132} &   \colhead{TRAPPIST-1}}
\startdata 
Source Counts  (ct) & 224  & 970 & 22,484 & 7,015 & 104 &  107 \\
Count Rate  (ct ks$^{-1}$) & 25$\pm$2  & 54$\pm$3 & 740$\pm$6  & 305$\pm$14 & 2.2$\pm$0.4  & 5.4$\pm$0.6 \\
Variable (Yes/No) & N & N & Y  & Y &  Y?  & (Y)  \\
f$_{X}$ (10$^{-14}$ ergs cm$^{-2}$ s$^{-1}$)  & 4.0$\pm$0.5 & 12.5$_{-1.5}^{+3.2} $ &134$\pm$2 &568$\pm$7  & 1.12$_{-0.27}^{+0.33}$ & $1.15_{-0.22}^{+0.23}$\\
$\log$ L$_{X}$ (erg s$^{-1}$)   & $26.07\pm0.06$ & $27.21\pm0.06$ & $27.52\pm0.06$ &  $27.78\pm0.06$ &  $26.33\pm0.06$  &  $26.33\pm0.06$\\
log L$_{X}$/L$_{bol}$   &  -5.95$\pm$0.06  &  -5.02$\pm$0.06 &   -4.26$\pm$0.06 &  -3.41$\pm$0.06 &  -4.89$\pm$0.06  &   -4.04$\pm$0.06  \\
kT$_{1}$ (keV)   &  0.23$\pm$0.02 & 0.12$_{-0.01}^{+0.04}$ & 0.12$_{-0.02}^{+0.01}$  & 0.15$\pm$0.02   & (0.086)  & (0.2) \\
VEM$_{1}$ (10$^{49}$ cm$^{3}$) & 0.18$\pm$0.02 & $0.83_{-0.03}^{+0.05}$ & $1.28_{-0.23}^{+0.18}$
& $1.83\pm0.02$ &  $0.38_{-0.13}^{+0.15}$    &  1.06$\pm$0.47   \\
kT$_{2}$ (keV)      &  \nodata & $0.62_{-0.06}^{+0.07}$ & $0.27_{-0.02}^{+0.01}$ &  0.34$\pm$0.02  & (0.86)  &  (0.8)  \\
VEM$_{2}$ (10$^{49}$ cm$^{3}$) & \nodata & $0.31_{-0.03}^{+0.01}$ &  $1.84_{-0.18}^{+0.14}$  & $3.32\pm0.04$  & $0.03\pm0.01$ &  $0.82\pm0.32$   \\
kT$_{3}$ (keV)         & \nodata & \nodata &$0.81_{-0.03}^{+0.05}$ & $0.75_{-0.03}^{+0.04}$   & \nodata  & \nodata\\
VEM$_{3}$ (10$^{49}$ cm$^{3}$) &\nodata &  \nodata & $0.65_{-0.09}^{+0.07}$ & $2.71\pm0.03$  & \nodata  & \nodata  \\
Red. $\chi^{2}$   & 1.15\tablenotemark{d}  & 1.32\tablenotemark{d} & 1.14\tablenotemark{d}  & 1.18\tablenotemark{e}  &  1.29\tablenotemark{f} & 1.29\tablenotemark{f}   \\
\enddata
\tablenotetext{a}{EPIC pn only}
\tablenotetext{b}{Fit performed using C-statistic, $\chi^2$ provided for reference only.}
\tablenotetext{c}{Fixed $N_H=1\times10^{19}\,$cm$^{-2}$ for the fit.}
\tablenotetext{d}{Abundances fixed to 0.4 solar}
\tablenotetext{e}{Adjusted abundances: Low-FIP (Fe, Mg): $0.47_{-0.06}^{+0.08}$, Mid-FIP (C, N, O, Si, S): $0.39_{-0.11}^{+0.09}$, High-FIP (Ne, Ar): $0.46_{-0.13}^{+0.09}$}
\tablenotetext{f}{Temperatures fixed. Abundances fixed to 0.4 solar.}
\end{deluxetable*}

\begin{longrotatetable}
\begin{deluxetable*}{ccccccccc}
\tabletypesize{\scriptsize}
\tablecaption{MUSCLES Chandra/ACIS M Dwarf Datasets and Measured Coronal
X-ray Properties\label{table5}}
\tablewidth{900pt}
\tablehead{
\colhead{ }  & \colhead{GJ 176} & \colhead{GJ 436} & \colhead{GJ 581} &
\colhead{GJ 581(flare)} & \colhead{GJ 667C}&
\colhead{GJ 876(low)}& \colhead{GJ 876(high)}& \colhead{GJ 1214} }
\startdata
Source Counts (ct)& 225& 281& 102& 250& 168& 121& 910 & \nodata \\
Count Rate (ct ks$^{-1}$)&7.7$\pm$0.5&2.5$\pm$0.2&2.5$\pm$0.3& 35.7$\pm$2.3& 
6.0$\pm$0.5& 12.1$\pm$1.1& 45.9$\pm$1.5& $\leq$0.14 (90$\%$ Conf.)  \\
Variable (Yes/No) & N & N & Y & Y & N & Y & Y & (Y)  \\
f$_{X}$ (10$^{-14}$ erg cm$^{-2}$ s$^{-1}$)&4.8$\pm$0.3&1.2$\pm$0.1&
1.8$\pm$0.2 & 21.5$\pm$1.4 & 3.9$\pm$0.3  &
9.1$\pm$0.8  & 30.5$\pm$1.0 & $\leq$0.11\tablenotemark{d}   \\
log L$_{X}$ (erg s$^{-1}$) &26.71$\pm$0.03 &26.14$\pm$0.04&25.93$\pm$0.05
 &27.01$\pm$0.06  &26.39$\pm$0.03   & 26.38$\pm$0.06  &
26.90$\pm$0.05  & $\leq$25.45   \\
log L$_{X}$/L$_{bol}$ &-5.42$\pm$0.04 &-5.84$\pm$0.04&-5.73$\pm$0.05 &
-4.65$\pm$0.06  & -5.35$\pm$0.04 &-5.32$\pm$0.06 &-4.80$\pm$0.06 &
$\leq$-5.7  \\
kT$_{1}$ (keV) & 0.31$\pm$0.02&0.39$\pm$0.03&0.26$\pm$0.02 &
0.99$\pm$0.08 & 0.41$\pm$0.03 & 0.80$\pm$0.14 & 0.68$\pm$0.04 & \nodata \\
VEM$_{1}$ (10$^{49}$ cm$^{3}$)&5.3$\pm$1.5&1.04$\pm$0.25 &1.00$\pm$0.30&
2.64$\pm$0.66& 1.81$\pm$0.54 &0.37$\pm$0.15& 6.5$\pm$1.2& \nodata  \\
kT$_{2}$ (keV) & \nodata & \nodata & \nodata & 0.27$\pm$0.06 & \nodata &
 0.14$\pm$0.04& 0.24$\pm$0.08  & \nodata \\
VEM$_{2}$ (10$^{49}$ cm$^{3}$) & \nodata &\nodata & \nodata  &
2.30$\pm$0.97 & \nodata& 1.3$\pm$1.6 & 1.8$\pm$1.2 & \nodata  \\
Red. $\chi^{2}$ & 0.88\tablenotemark{a}& 0.7\tablenotemark{b} &
1.13\tablenotemark{c} & 1.23\tablenotemark{d} & 1.63\tablenotemark{e} & 
 0.98\tablenotemark{d} & 1.03\tablenotemark{f} & \nodata \\
\enddata
\tablenotetext{a}{Adjusted abundances: Fe=0.41$\pm$0.18}
\tablenotetext{b}{Adjusted abundances: Fe=0.17$\pm$0.11}
\tablenotetext{c}{Adopted sub-solar abundances: Fe=0.15}
\tablenotetext{d}{Adopted solar abundances: Fe=1.0}
\tablenotetext{e}{Adjusted abundances: Fe=0.19$\pm$0.12}
\tablenotetext{f}{Adopted sub-solar abundances: Fe=0.195;
measured Fe=0.17$\pm$0.04}
\end{deluxetable*}
\end{longrotatetable}

\begin{longrotatetable}
\begin{deluxetable*}{ccccccccc}
\tabletypesize{\footnotesize}
\tablecaption{Mega-MUSCLES M Dwarf Datasets and Measured Coronal
X-ray Properties\label{table6}}
\tablewidth{950pt}
\tablehead{
\colhead{ }  & \colhead{GJ 15A} & \colhead{GJ 15B} & \colhead{GJ 163} &
\colhead{GJ 849} & \colhead{GJ 3843} & \colhead{GJ 699 (all)} &  \colhead{GJ 699 (quiet)} &
\colhead{LHS2686} }
\startdata
Instrument&ACIS-S3&ACIS-S3&ACIS-S3&ACIS-S3&ACIS-S3&ACIS-S3&
ACIS-S3&ACIS-S3  \\
Source Counts (ct)& 175& 100& 41& 77 &\nodata & 137 & 94  & 469 \\
Count Rate (ct ks$^{-1}$)&7.3$\pm$0.6&4.2$\pm$0.4&1.4$\pm$0.3& 2.8$\pm$0.3& 
$\leq$0.86 &  5.2$\pm$0.5& 4.0$\pm$0.4 &  $17.6\pm0.8$  \\
Variable (Yes/No) & N & Y & N & N & \nodata & Y & N &   Y  \\
f$_{X}$ (10$^{-14}$ erg cm$^{-2}$ s$^{-1}$)&8.5$\pm$0.7&5.1$\pm$0.5&
1.97$\pm$0.34 & 4.74$\pm$0.57 & $\leq$0.81`\tablenotemark{a} &
6.30$\pm$0.55  & 4.83$\pm$0.52 &15.3$\pm$0.7      \\
log L$_{X}$ (erg s$^{-1}$) &26.11$\pm$0.04 &25.89$\pm$0.08 &26.73$\pm$0.06
 &26.64$\pm$0.06 & $\leq$25.28  &25.40$\pm$0.06 & 25.29$\pm$0.06 & 
 27.44$\pm$0.06  \\
log L$_{X}$/L$_{bol}$ &-5.82$\pm$0.04 &-5.27$\pm$0.08 &-5.19$\pm$0.08 &
-5.40$\pm$0.06 &  $\leq$-5.15 &-5.71$\pm$0.06& -5.83$\pm$0.06 &
-3.60$\pm$0.06  \\
kT$_{1}$ (keV) & 0.45$\pm$0.07&0.56$\pm$0.12&0.51$\pm$0.22 &
0.32$\pm$0.05 &  \nodata & 0.54$\pm$0.11 &
0.49$\pm$0.12& 0.90$\pm$0.05  \\
VEM$_{1}$ (10$^{49}$ cm$^{3}$)&1.22$\pm$0.33& 0.68$\pm$0.18 & 4.9$\pm$3.8&
    4.4$\pm$1.9 &  \nodata& 0.23$\pm$0.07 & 0.11$\pm$0.08 &12.1$\pm$3.0  \\
Red. $\chi^{2}$ & 1.12\tablenotemark{b}& 1.25\tablenotemark{c} &
0.87\tablenotemark{b} & 1.35\tablenotemark{b} & \nodata & 
1.03\tablenotemark{d} &  1.20\tablenotemark{e} & 1.02\tablenotemark{f}  \\
\enddata
\tablenotetext{a}{Adopted kT=0.43 keV and solar abundances: Fe=1.0 }
\tablenotetext{b}{Adopted sub-solar abundances: Fe=0.195}
\tablenotetext{c}{Adopted sub-solar abundances: Fe=0.195,Using L$_{bol}$ = 1.44 $\times$ 10$^{31}$ erg s$^{-1}$}
\tablenotetext{d}{Adjusted abundances: Fe=0.19$\pm$0.12; Adopted N$_H$ = 5 $\times$ 10$^{18}$ cm$^{-2}$}
\tablenotetext{e}{Adjusted abundances: Fe=0.36$\pm$0.23; Adopted N$_H$ = 5 $\times$ 10$^{18}$ cm$^{-2}$}
\tablenotetext{f}{Adjusteded sub-solar abundances: Fe=0.61$\pm$0.12}
\end{deluxetable*}
\end{longrotatetable}

\begin{figure*}[h]
\includegraphics[angle=90,scale=.70]{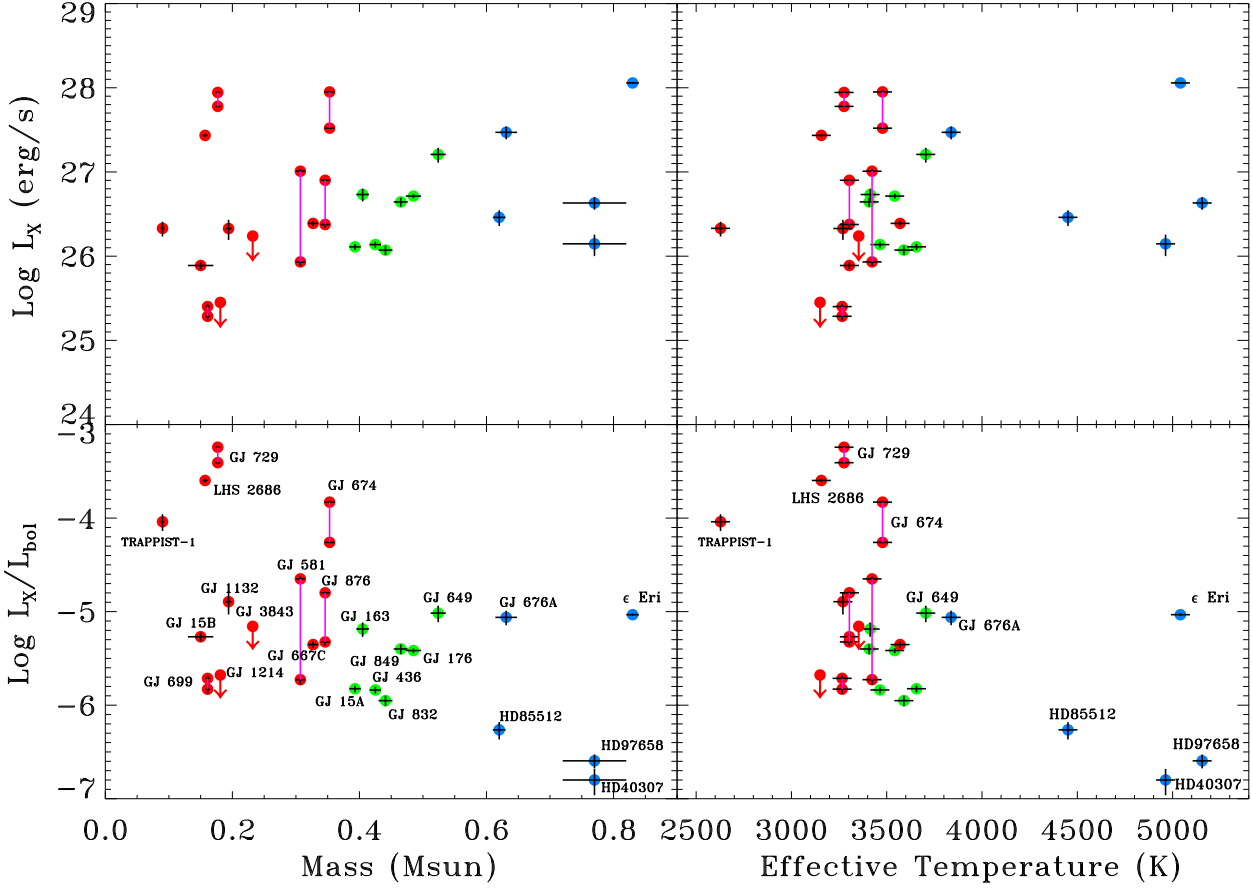}
\caption{X-ray (0.3 - 10.0 keV) luminosity and X-ray-to-bolometric luminosity ratio as a function 
of stellar mass and effective temperature. Observations of the same star in quiescent and flaring states 
are connected by vertical lines. Color coding: blue - M$\geq$ 0.6 M$_{\odot}$; 
green - 0.35$\leq$ M $\leq$ 0.6 M$_{\odot}$;  red M $\leq$ 0.35 M$_{\odot}$ . Error bars are plotted in black 
with some being smaller than the plotting symbols. 
\label{figLxMT}}
\end{figure*}

\section{Coronal Physical Properties}\label{sec:coronal-props}

The derived X-ray luminosities and the X-ray-to-bolometric luminosity ratios 
are the primary observables that can be compared to the stellar physical 
properties discussed in Sect. \ref{sec:sample}. 

\begin{figure}
\includegraphics[angle=90,scale=.73]{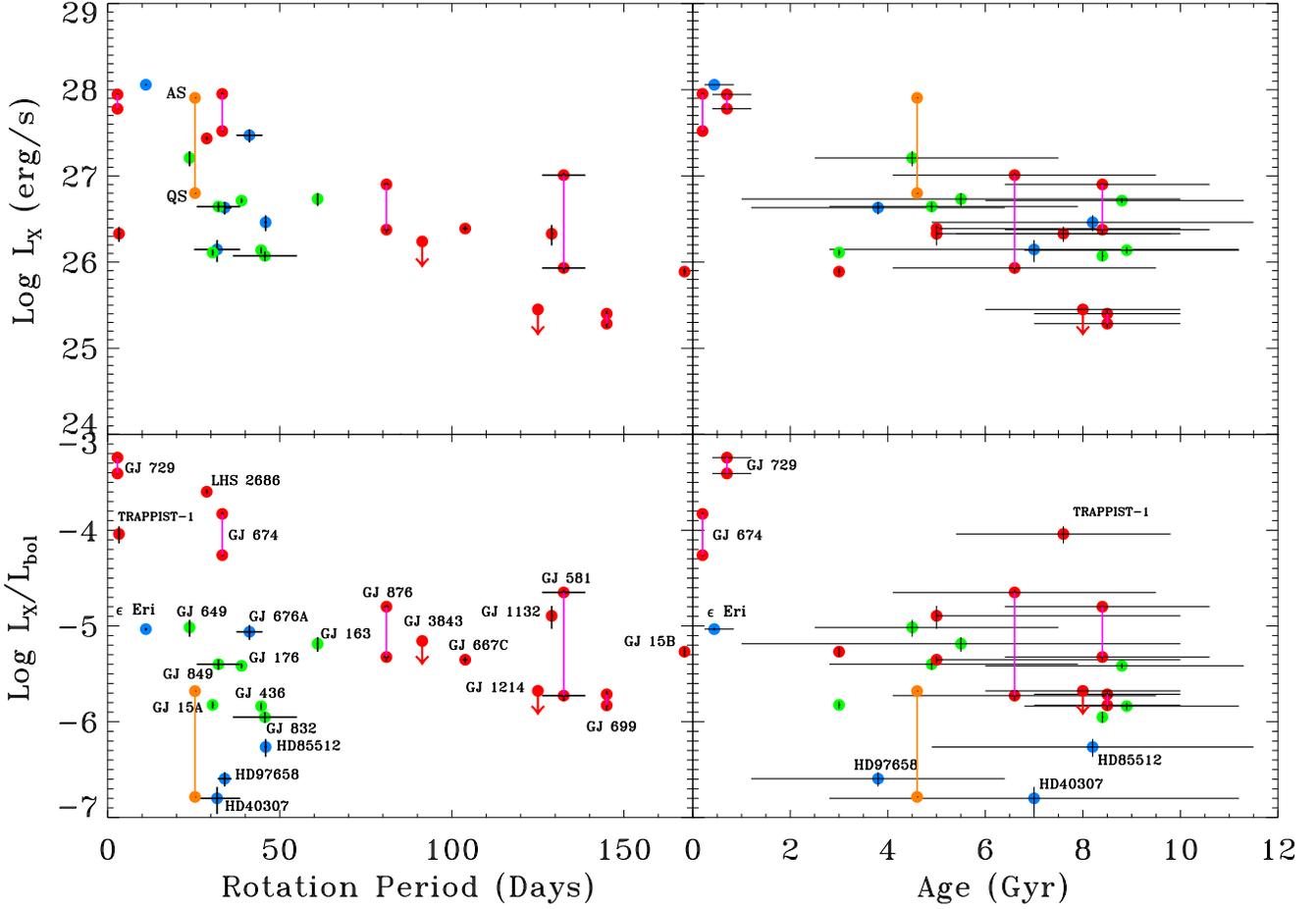}
\caption{X-ray luminosity and X-ray-to-bolometric luminosity ratio as a function 
of stellar rotation period (left panels) and age (right panels). For comparison, 
the equivalent numbers for the Quiet Sun (QS) and Active Sun (AS) are shown 
as orange symbols.
Most stars in the sample are older than the Sun (4.5 Gyr). 
Three stars are younger than 1 Gyr. For older stars 
the age uncertainties are large. Color coding as in Fig. \ref{figLxMT}.  \label{figLxPA}}
\end{figure}

Figures \ref{figLxMT} and \ref{figLxPA} show the distributions of the soft-X-ray 
luminosity and X-ray-to-bolometric luminosity ratio with stellar mass, effective 
temperature, rotational period and age, respectively. Where feasible, identification of 
the individual stars in the sample is provided, because many exoplanet researchers will 
be interested about particular stars and how they fall within the overall sample.

Generally the X-ray luminosities of the sample lie between 10$^{26}$ and 10$^{28}$
erg s$^{-1}$, with only GJ 699 being detected at a significantly lower luminosity aided 
by its proximity at a distance of only 1.8 pc .  The mean X-ray luminosities 
for the more massive (0.35 M$_{\odot}$ $\leq$ M $\leq$ 0.6 M$_{\odot}$) M dwarfs is 
$\sim$5 $\times$ 10$^{26}$ erg s$^{-1}$ compared to $\sim$2 $\times$ 10$^{26}$ erg s$^{-1}$ 
for fully-convective stars older than 1 Gyr. Younger, fully-convective M dwarfs have X-ray 
luminosities between 3 and 6 $\times$ 10$^{27}$ erg s$^{-1}$.
For comparison, the corresponding X-ray 
luminosities from the Quiet and Active Sun are 6.31 $\times$ 10$^{26}$ and 
8.04 $\times$ 10$^{27}$ erg s$^{-1}$ respectively and  the L$_{X}$/L$_{bol}$ ratios 
are 1.64 $\times$ 10$^{-7}$ and 2.09 $\times$ 10$^{-6}$ \citep{linsky20}.
This sensitivity level is imposed by the capabilities of existing X-ray observatories, 
but, nevertheless the vast majority of the sample provided secure X-ray detections 
and multiple examples of significant coronal variability. When considered in terms 
of X-ray to bolometric luminosity, the major increase in bolometric luminosity from 
late-M to early-K stars contributes to the trend  where X-ray emission 
from non-saturated coronae becomes a larger component of the stellar emission 
for stars of lower mass  \citep{fleming95}.

\section{Coronal Variability}\label{sec:variability}
The X-ray emission from K and M dwarfs typically shows considerable variability that results in 
major changes of the radiation field impacting exoplanets. This variability is not confined to the 
most active, rapidly-rotating stars but extends to the oldest, least active stars as well. We have 
examined all our data sets for short-term variability and compared our results with those published 
by others for long-term variability. For discussion we divide our sample into three mass ranges.  

During the MUSCLES/Mega-MUSCLES 
X-ray observations flare outbursts are seen from seven of ten fully-convective stars. 
Younger fully-convective stars all showed multiple flare outbursts. However, in all cases only a small 
portion of the time sampled contained flares. For example, even for GJ 876A (see Fig. \ref{figGJ876_GJ581}) 
the X-ray flux is only above three times the quiescent level for 31\% of the observation.
For seven fully-convective stars stars older than 1 Gyr six flare outbursts are seen over 226 ks.
In contrast, the three older K dwarfs show zero discernible flares over 78 ks and seven 
more massive M dwarfs show zero discernible flares over 244 ks.

\subsection{K - M0 Dwarfs -- M$\geq$0.6 M$_\odot$}\label{subsec:Kstars} 

Our sample contains 5 stars in this mass range with only $\epsilon$ Eri considered to 
be young.  None of the stars shows X-ray flaring in dedicated MUSCLES X-ray observations but archival 
observations show the presence of some long-term variability.

$\epsilon$ Eri (HD 22409, GJ 144; K2 V) : This star is the youngest K dwarf  in our sample 
and shows bright X-ray emission with a luminosity exceeding 10$^{28}$ erg s$^{-1}$.
The MUSCLES XMM-Newton EPIC pn observation of $\epsilon$ Eri  does not show significant 
coronal variability. \citet{Loyd_etal18} observed an FUV flare with HST immediately preceding 
the EPIC pn observations and an X-ray flare decay was possibly seen by the EPIC MOS detectors 
but with large error bars.  Two small FUV enhancements during the EPIC pn observation 
produced no detectable corresponding X-ray response. More extensive 
monitoring of its X-ray emission with a further 7 observations over the following 3.5 years 
shows quiescent flux levels up to twice that seen in 2015 and additional short-term flaring 
to over 2.1 $\times$ 10$^{-11}$ erg cm$^{-2}$ s$^{-1}$  \citep{coffaro20}. These variations are 
reasonably consistent with a likely 2.9 year period activity cycle seen in Ca II HK detected by 
\citet{metcalfe13}.

The data for HD 40307 (GJ 2406; K2.5 V), HD 85512 (GJ 370, K6 V), and HD 97658 (K1 V), and 
GJ 676A (M0 V) show no statistically significant variability,
However, for HD 97658 there is a suggestion of variability between the 2015-10-17 
Chandra observation (count rate 1.31$\pm$0.35$\times$10$^{-3}$ ct s$^{-1}$; 3.7$\sigma$), 
which showed twice the count rate seen in the two later observations (2015-12-11: 
count rate 0.73$\pm$0.22$\times$10$^{-3}$ ct s$^{-1}$; 3.3$\sigma$, and 2016-03-05: 
count rate 0.61$\pm$0.21$\times$10$^{-3}$ ct s$^{-1}$; 2.9$\sigma$).

\subsection{Early-Mid M Dwarfs (M1-3.5) with M$\geq$0.35 M$_\odot$ }\label{subsec:earlyMstars} 

Seven early-mid M dwarf stars fall into this mass  category and none of them show significant X-ray 
flaring. While GJ 667C (HD156384C) is usually quoted to have an M1.5 V spectral type, its mass 
(0.327 M$_\odot$) and effective temperature (3443 K) indicate that it is more likely to be a 
fully convective star. Therefore, GJ 667C is discussed in Sect. \ref{subsec:midMstars}

\begin{figure}[t]
\includegraphics[angle=90,scale=.53]{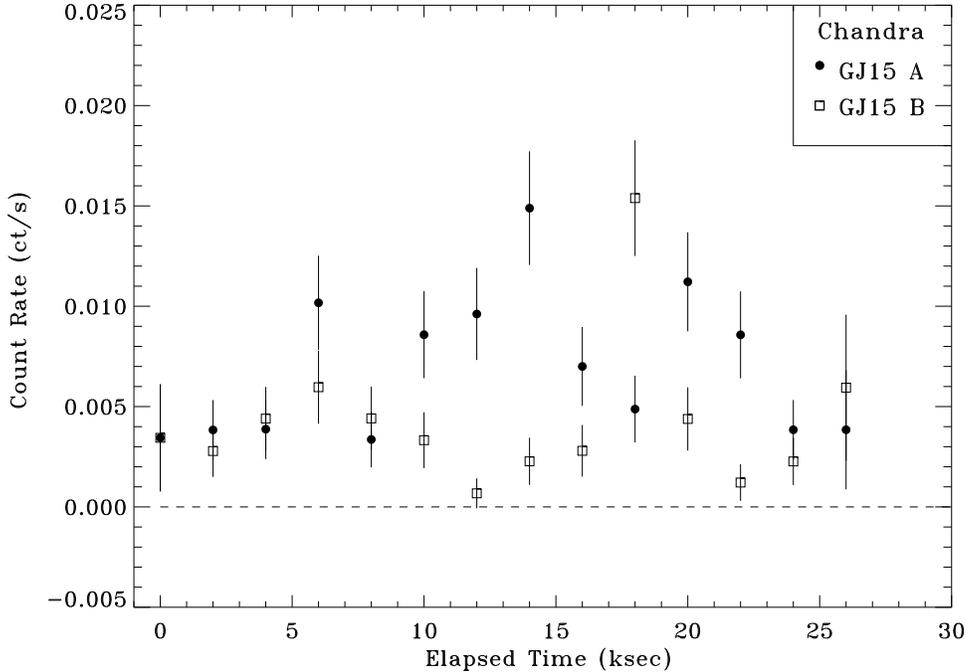}
\caption{Chandra ACIS-S3 lightcurve of GJ 15 A and B using 2 kilosecond
sampling bins. 1$\sigma$ error bars are plotted.  \label{figGJ15}}
\end{figure}

GJ 15A (M2 V) and GJ 15B (M4 V) :  The two stars in this wide binary were observed 
in the same ACIS-S3 observations and show independent patterns of variability 
(see Fig. \ref{figGJ15}). Both stars are close to the boundary of significant variability with GJ 15B 
above the threshold (glvary VARINDEX = 8) and GJ 15A  below (glvary VARINDEX = 1).

GJ 649 (M1 V), GJ 832 (HD 204961; M1.5 V),  
GJ 176 (HD 285968, M2 V, Fig. \ref{figGJ176}), GJ 436 (M3 V, Fig.  \ref{figGJ436}), 
GJ 163 (M3.5 V), GJ 849 (M3.5 V, Fig.  \ref{figGJ849}): 
All showed no statistically significant variability and the light curves for some of these stars 
are shown at the end of this paper in Appendix A. The absence of detectable X-ray flaring 
on these stars is somewhat surprising because FUV flares were clearly seen in simultaneous 
HST observations of GJ 176 (see Fig. \ref{figGJ163}). Low count rates for many of these stars 
necessitate light curves with large temporal bins that may smooth out small scale flare signals.

\subsection{Fully Convective M Dwarfs (M3-8) with M$\leq$0.35 M$_\odot$ }\label{subsec:midMstars}

There are ten fully convective stars in the sample and all but one of those that are detected 
(plus GJ 15B) show short-term variability.

\begin{figure*}
\includegraphics[angle=90,scale=.65]{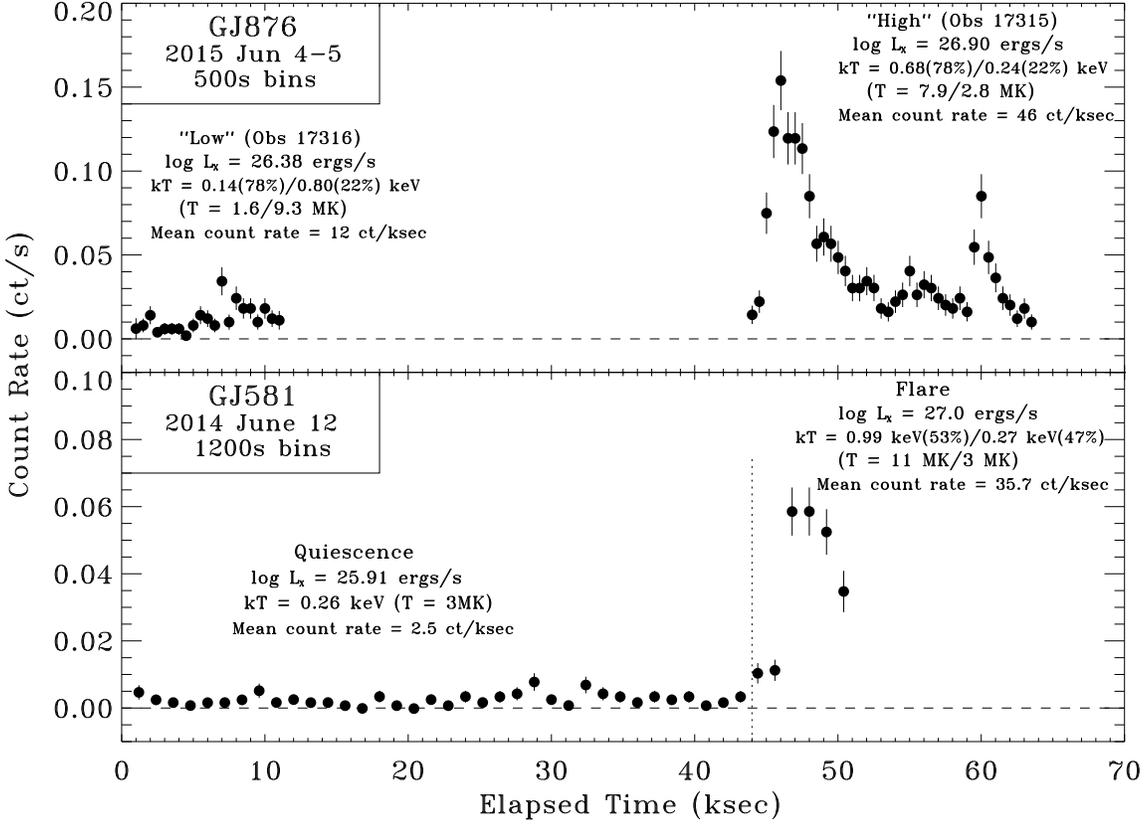}
\caption{Chandra ACIS-S3 light-curves of GJ 876A (M4 V; P$_{rot}$ = 81 d) and 
GJ 581 (M3 V; P$_{rot}$ = 132 d). Both stars have masses between 0.30 and 0.35 M$_\odot$ 
and are thus in the fully convective mass range.
Despite having long rotation periods, both stars show dramatic coronal flaring.  1$\sigma$ error bars are plotted. 
The vertical dotted line in the GJ 581 figure marks the time split 
between the quiescent and flaring time intervals -- the quiescent exposure totals 
40.6 ksec, while the flare is observed for 7.0 ksec .  \label{figGJ876_GJ581}}
\end{figure*}

GJ 876A (M4 V) : 
This star is a fairly slowly rotating mid-M dwarf but 
shows significant FUV and X-ray flaring activity (\citet{france12}, \citet{france16}).
\citet{rivera05}  measured an initial photometric rotation period of 96.7 days 
and \citet{diez_alonso19} more recently produced a revised period of
81.0 days. This rotational period indicates that the rotation is still 
slowing and significant rotational evolution is yet to occur. 
The Chandra X-ray light curve is shown  in Fig. \ref{figGJ876_GJ581}. Dramatic soft X-ray flaring 
is present throughout the observations. The data were obtained in two segments with an 
initial "low" state observation and then a longer "high" state observation containing 
multiple flares that occur throughout the 20 ks observation. 
The minimum 0.3-10.0 keV luminosity at the start of the first segment 
was 1.1 $\times$ 10$^{26}$ erg s$^{-1}$, compared to a maximum flare peak 
luminosity of 2.6 $\times$ 10$^{27}$ erg s$^{-1}$, which is an increase by a factor of 23.
The total energies over the same energy range in the initial large and smaller final flares 
were 1.24 $\times$ 10$^{31}$ ergs and 2.72 $\times$ 10$^{30}$ ergs, respectively.  
The total integrated energy during the "High" observation was 1.54 $\times$ 10$^{31}$ ergs.
For direct comparison to solar flares, the soft X-ray flux in the GOES 1-8 \AA\ (1.55-12.4 keV)
passband was estimated (see \citet{Youngblood_etal17}). During the peak of the largest flare  
 a 1T XSPEC coronal temperature of 0.73 keV was measured. At this time 11.2\% of the 
 Chandra flux fell in the GOES SXR band, and thus implies a "GOES" flux of 
 8.2$^{+0.8}_{-1.0}$ $\times$ 10$^{-14}$ erg cm$^{-2}$ s$^{-1}$ and effects equivalent to a 
 M9.5 (error range M7.8-X1.1) solar flare in the habitable zone of GJ876A.
\citet{Youngblood_etal17} provide a detailed discussion of the potential impact of 
flare radiation and particle fluxes in this star's habitable zone.

GJ 581 (M3 V) : Despite showing slow rotation with  period estimates of 132.5 $\pm$6.3 days 
\citep{mascareno15} and 130$\pm$2 days \citep{robertson14a}, the Chandra observation of 
GJ 581 showed a major flare (see Fig. \ref{figGJ876_GJ581}). The flare peak was at a 
flux level more than 20 times quiescence. During the flare half the emission was from plasma 
with a characteristic temperature of 1.0 keV (11 MK), while the other half was near the 
quiescent temperature of 0.27 keV (3 MK) but with twice the quiescent volume emission 
measure. The flare is seen for 7 ks until the end of the observation, when the emission is still 
far above the quiescent level.

\begin{figure}[h]
\includegraphics[angle=90,scale=.70]{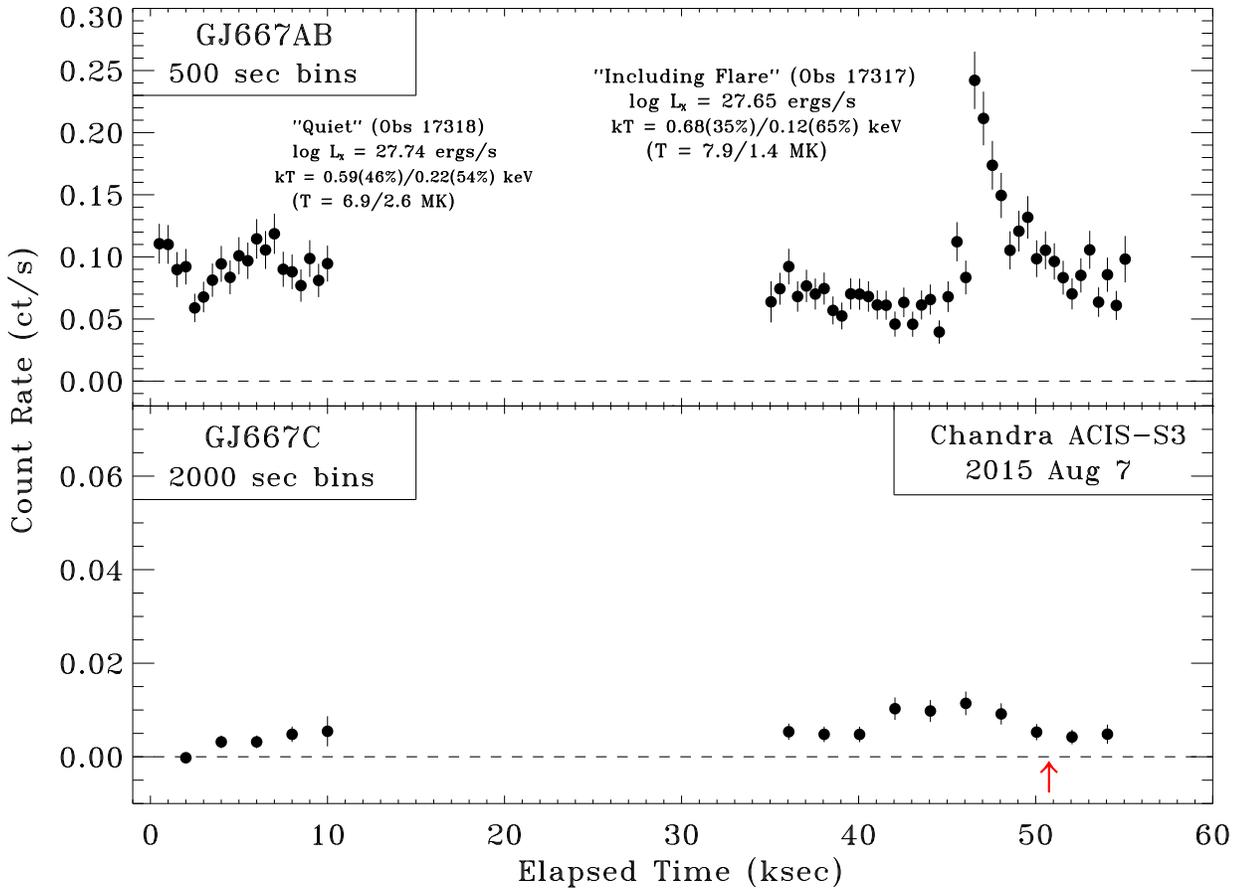}
\caption{Chandra ACIS-S3 light curves of the K3+K5 binary GJ 667AB (upper panel) 
using 500 second binning and GJ 667C (lower panel) using 2 kilosecond sampling bins. 
1$\sigma$ error bars are plotted. 
 Note that the GJ 667C y-axis covers only one quarter the count rate range of the GJ667AB plot.
GJ 667C is only a weak source and no statistically significant variability is detected.  
A red upward arrow marks the time when an FUV flare was detected 
\citep{Loyd_etal18}.   \label{figGJ667}}
\end{figure}

GJ 667C (HD156384C, M1.5 V): This M dwarf is the tertiary component in a hierarchical 
triple star system with the more massive A (K3) and B (K5) components forming a 42 yr period binary 
\citep{malkov12}.  Fig. \ref{figGJ667}  shows the X-ray light curves for the C and combined AB components.
GJ 667C is a weak non-variable source but the AB pair shows variable X-ray emission with 
one obvious flare. For GJ 667C an FUV flare was seen during simultaneous HST observations \citep{Loyd_etal18} 
but no X-ray counterpart is visible. The A and B components are barely separated by ACIS with GJ 667A 
being the brighter X-ray source contributing $\sim$70-75\% of the X-ray emission of the 
K dwarf pair. The mean X-ray luminosity of GJ 667AB is 4.86 $\times$ 10$^{27}$ erg s$^{-1}$ 
with almost equal fluxes measured from the two Chandra observations, The X-ray to bolometric 
luminosity ratio for the combined AB signal is log L$_{X}$/L$_{bol}$ $\sim$ -5.5.

\begin{figure*}
\includegraphics[angle=90,scale=.65]{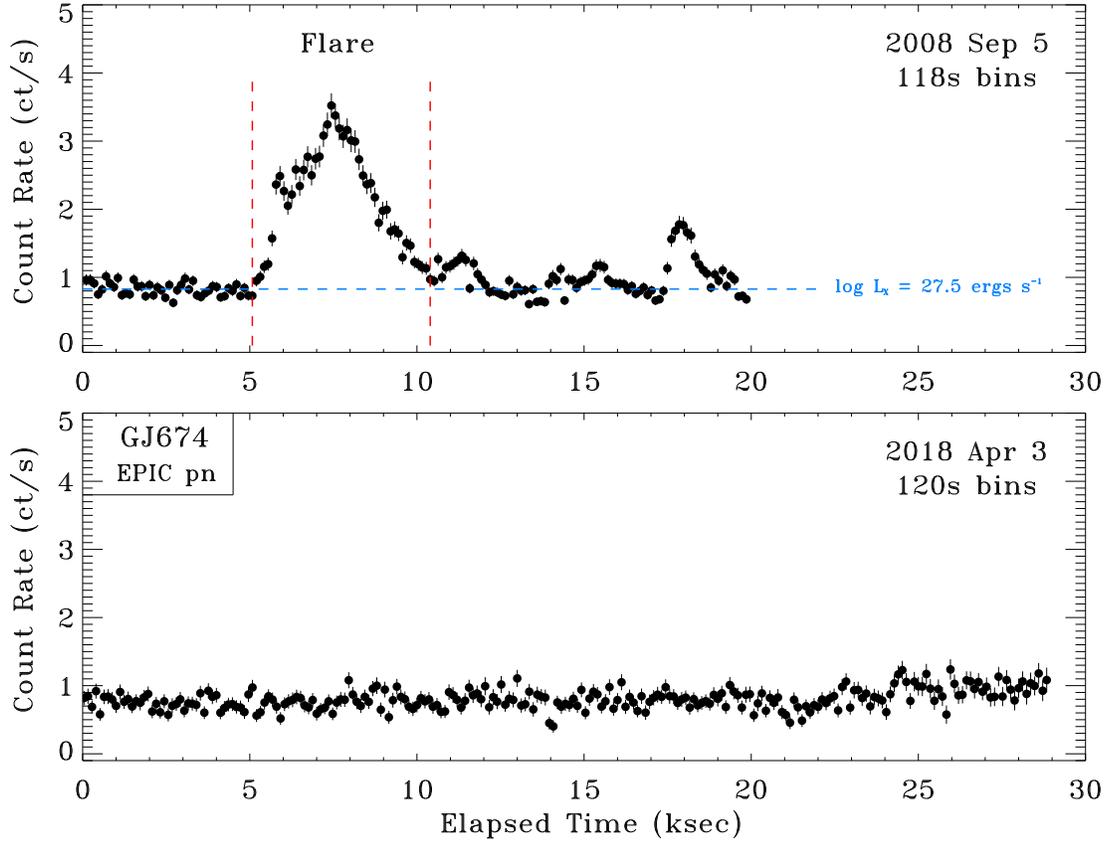}
\caption{XMM-Newton EPIC pn light-curves of the young star GJ 674 (M3V; P$_{rot}$ = 33 d).
 This star has a mass of 0.35 M$_\odot$ and thus lies right at the fully convective boundary.
GJ 674 showed significant X-ray flaring in 2008 but little variability in 2018. However, 
immediately after the 2018 XMM-Newton observation ended an extremely large, high 
temperature FUV flare was seen by HST COS \citep{froning19}. 
 Dashed vertical lines delineate the large 2008 flare. 1$\sigma$ error bars are plotted. \label{figGJ674}}
\end{figure*}

GJ 674 (M3 V): Two XMM-Newton datasets exist for GJ 674. During our 2018 observation 
some weak variability was seen  but the observations ended at 14:40 on 2018 Apr 3 
just minutes before the onset of the giant FUV flare observed by HST \citep{froning19}.
However, observation 0551020101 (PI Schmitt) on 2008 Sep 5 contained one
large and two small X-ray flares \citep{poppenhaeger10}. The EPIC pn lightcurves are 
shown in Fig. \ref{figGJ674}. The quiescent count rates are almost identical in the 
two observations. The average X-ray luminosity during the larger 2008 flare is 
8.9 $\times$ 10$^{27}$ erg s$^{-1}$ (log L$_X$ = 27.95 erg s$^{-1}$) and the peak 
luminosity is 1.5 $\times$ 10$^{28}$ erg s$^{-1}$ (log L$_X$ = 28.17 erg s$^{-1}$).
The duration of the flare was 5.3 ks. 
The integrated net flare energy (i.e. after subtraction of the quiescent emission) is 
3.0 $\times$ 10$^{31}$ ergs.

\begin{figure*}
\includegraphics[angle=90,scale=.65]{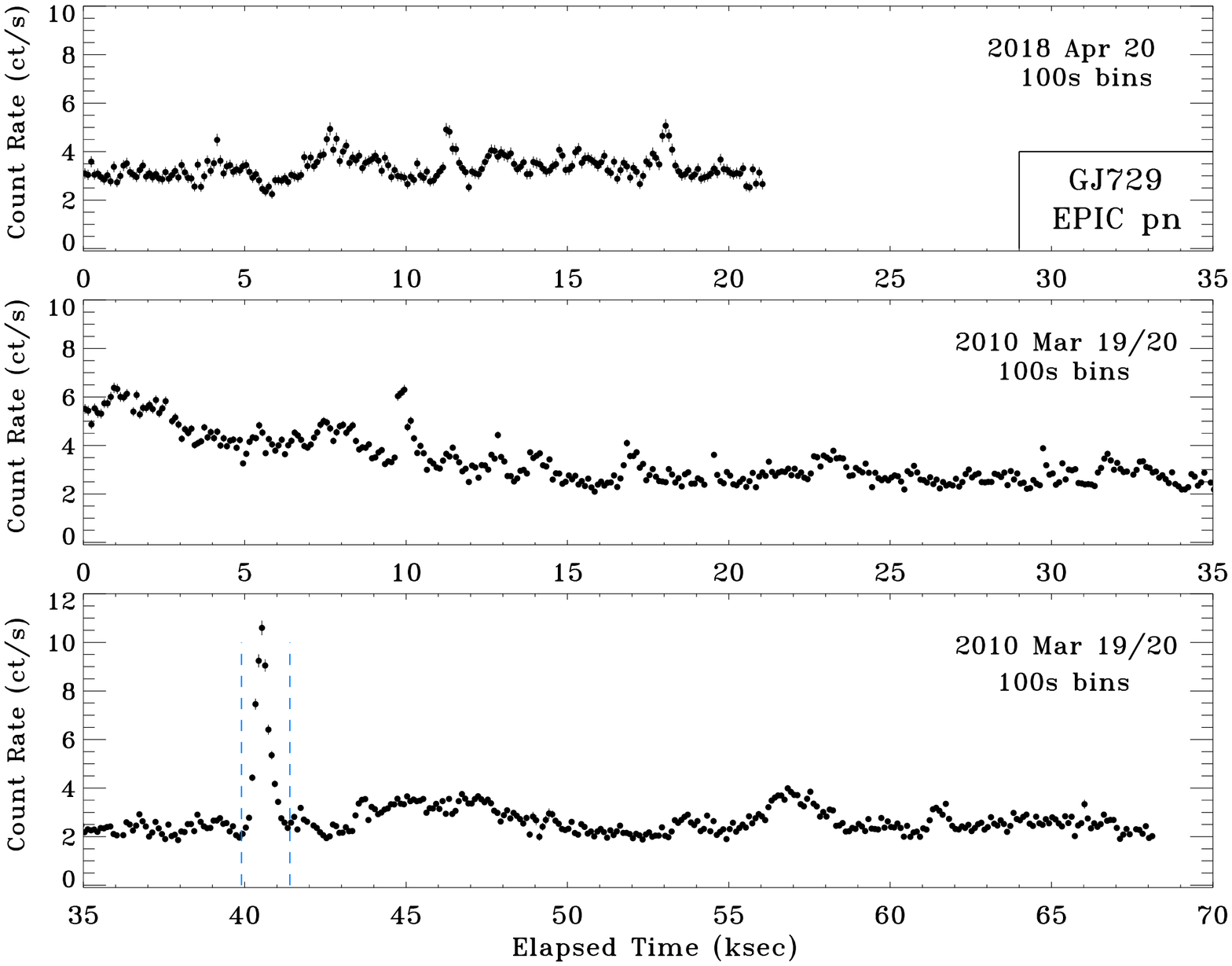}
\caption{XMM-Newton EPIC pn light-curves of the young star GJ 729 
(M4V; P$_{rot}$ = 2.85 d) 
from observations in 2018 and 2010. GJ 729 shows persistent X-ray flaring.  
Dashed vertical lines delineate the 2010 flare discussed in the text. 
1$\sigma$ error bars are plotted. \label{figGJ729}}
\end{figure*}

GJ 729 (Ross 154; M4 V): This star is the youngest fully-convective star in the sample 
with a rotation period of 2.848 days \citep{newton16} and an age of less than 
1 Gyr based on its membership in the Castor Moving Group \citep{allen98}.
Persistent X-ray variability is seen in  XMM-Newton observations (see Fig. \ref{figGJ729}),
 including our Mega-MUSCLES observation in 2018 and an earlier 68 ksec observation 
in 2010 (ObsID: 0601950101, PI Wargelin). The largest 2010 flare has 
 average X-ray luminosity of 8.8 $\times$ 10$^{27}$ erg s$^{-1}$ 
 (log L$_X$ = 27.94 erg s$^{-1}$) and the peak 
luminosity is 1.92 $\times$ 10$^{28}$ erg s$^{-1}$ (log L$_X$ = 28.28 erg s$^{-1}$).
The integrated net flare energy (i.e. after subtraction of the quiescent emission) is 
7.26 $\times$ 10$^{30}$ ergs, which is roughly twice the quiescent emission 
emitted during the 1.5 ks flare duration.

GJ 729 has a history of dramatic X-ray flaring, with very large flares seen in 
previous Chandra ACIS and HRC observations \citep{wargelin08}. During the 2002 
Sep 9 ACIS-S observation the X-ray luminosity rose from a quiescent level of
 9 $\times$ 10$^{27}$ erg s$^{-1}$ to a peak of 1.8 $\times$ 10$^{30}$ erg s$^{-1}$.
The flare peak showed a dominant  temperature component at  $\sim$3 keV 
(35 MK).

\begin{figure*}[h]
\includegraphics[angle=90,scale=.70]{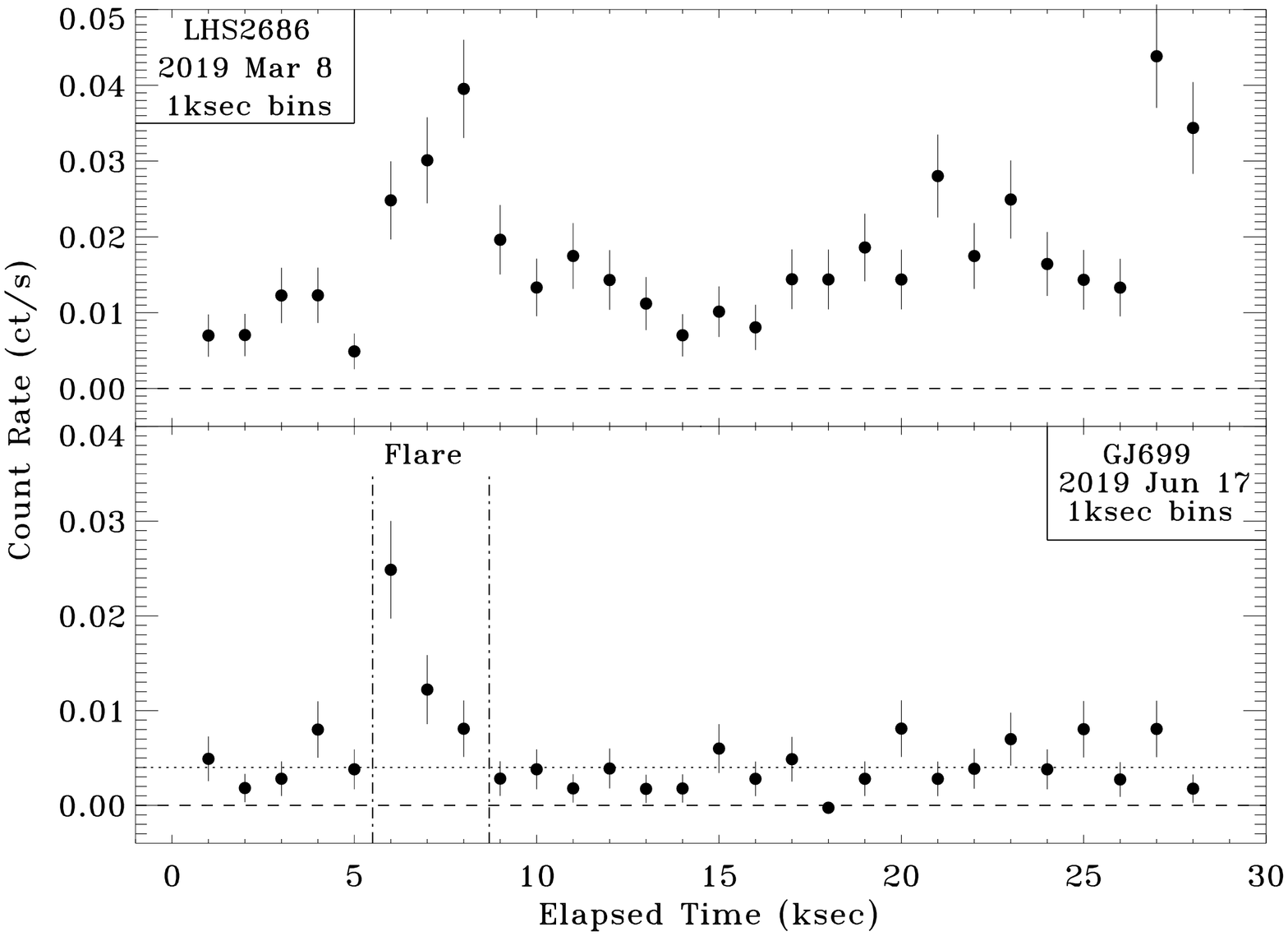}
\caption{Chandra ACIS-S3 light curves of LHS 2686 (M4.5 V; P$_{rot}$ = 29 d) 
and GJ 699 (M4 V; P$_{rot}$ = 145 d). These stars are likely to be fully convective 
having masses near 0.16 M$_\odot$. While GJ 699 is a slow rotator and LHS2686 
is a much faster rotator, both stars show significant flaring. 
Both plots are for 1 kilosecond sampling bins. 1$\sigma$ error bars are plotted. 
 \label{figLHS2686_GJ699}}
\end{figure*}

LHS 2686 (M4.5 V) : The age of LHS 2686 is unknown but its Chandra 
light curve (see Fig. \ref{figLHS2686_GJ699}) is highly variable with at least 2 flares
present and a glvary VARINDEX of 8.  LHS 2686 is likely to be young (age $\leq$ 1 Gyr) 
based on this level of flaring, its strong H$\alpha$ emission, and its high X-ray luminosity 
and coronal temperature.

GJ 699 (Barnard's Star; M4 V) : Barnard's Star is a prototypical example of an 
old inactive M dwarf star with a rotation period of 145 days 
and its proximity to the Sun allows very sensitive 
measurements of stellar activity. GJ 699 has the lowest quiescent X-ray luminosity 
(log L$_X$ = 25.29 erg s$^{-1}$) of all the stars in our sample 
(see Table \ref{table6} and Fig. \ref{figLxMT}) 
and has the lowest M dwarf L$_X$/L$_{bol}$ ratio. Despite this low activity level,
our Chandra data contain a significant X-ray flare and COS spectra also 
show dramatic FUV flares \citep{france20}. The Chandra ACIS light curve is shown 
in Fig. \ref{figLHS2686_GJ699} and these data produce a glvary VARINDEX of 7. 
At the peak of the flare the X-ray flux is 2.5 $\times$ 10$^{-13}$ erg cm$^{2}$ s$^{-1}$ 
(corresponding to log L$_X$ = 26.00 erg s$^{-1}$), 
which is just over 5 times larger than the quiescent flux. The integrated flare energy, 
after subtraction of the quiescent emission, is 1.6 $\times$ 10$^{29}$ ergs and 
corresponds to the equivalent of a solar X2-3 flare in the habitable zone of 
Barnard's Star. Flares of this size occur roughly once per month on the Sun. 
During a total monitoring duration of 52 ksec (0.6 days), the MegaMUSCLES 
non-simultaneous COS and Chandra observations detected 3 major flares.
\citet{france20} consider the potential CME particle fluxes that might be associated with 
these flares and, based on exoplanet atmospheric modeling, conclude that the
quiescent radiation field would drive minimal atmospheric erosion but that flare 
conditions would produce significant erosion from any habitable zone planets.

GJ 1132 (M4 V): This weak source possibly shows a rise in the count rate towards the 
end of the observation, but the observation suffers from high background radiation levels.

TRAPPIST-1 (M8 V) :
During the 2018 December 18 XMM-Newton observation of TRAPPIST-1 only weak 
X-ray emission was detected at a count rate of 0.005 ct s$^{-1}$ , corresponding to  
log L$_X$ =26.33 ergs s$^{-1}$. Even long XMM-Newton observations have recorded too few events to 
allow detailed variability studies and most observations overlap in estimated X-ray flux and luminosity. 
In earlier observations with the same instruments \citet{wheatley17} detected variable emission but,
within the error bars, the source properties overlap with those from our 2018 observation.

\begin{figure}[b]
\includegraphics[trim= 37mm 0mm 40mm 0mm, angle=-90,scale=.65]{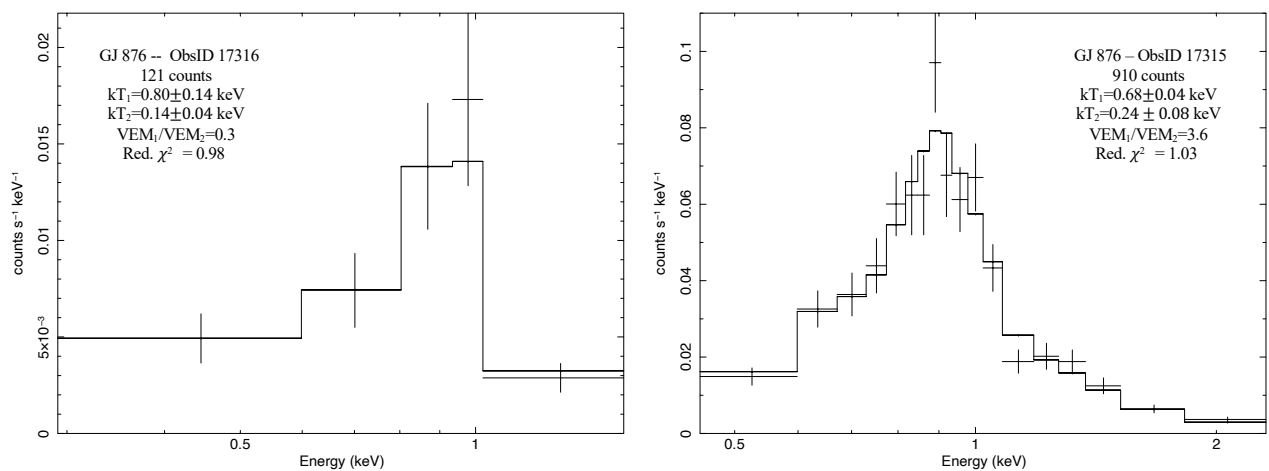}
\caption{Examples of XSPEC spectral fitting to Chandra CCD-resolution spectra of GJ 876. 
These figures demonstrate the fitting framework within XSPEC.
 The spectral data with associated error bars are overlaid on the final model shown as a histogram.
Right panel: a two-temperature (2-T) fit to the high activity (ObsID 17315) spectrum grouped 
with 35 events per channel.
Left panel: a 2-T fit to the low activity (ObsID 17316) spectrum, recorded 12 hours earlier, 
grouped with 15 events per channel.
\label{spectralfits}}
\end{figure}

\section{Results and Discussion}\label{sec:discuss}

\subsection{Coronal Thermal Structure}

\begin{figure}
\includegraphics[angle=90,scale=.65]{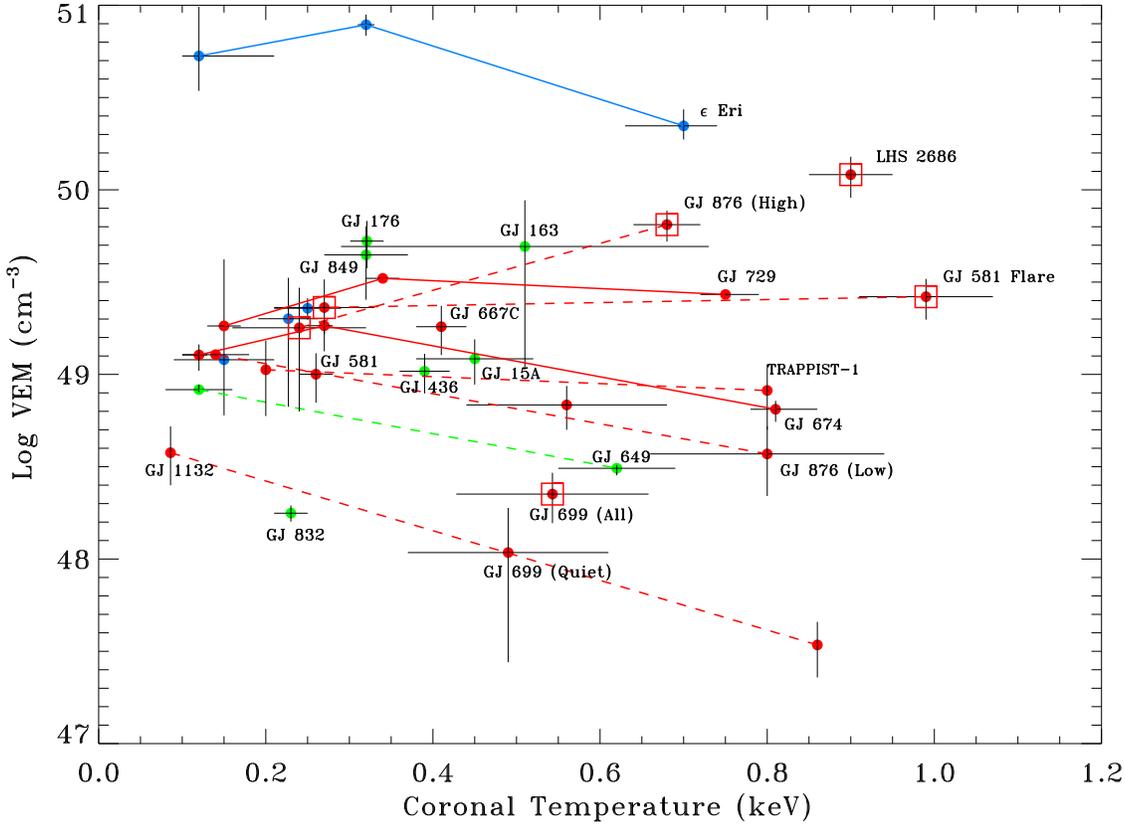}
\caption{Volume emission measure as a function of coronal temperature. The 
highest volume emission measure marks the dominant coronal thermal component 
for each star.  The coronal temperatures for GJ 1132 and TRAPPIST-1 are  assumed values and 
not measured values from the spectral fits.
Dashed lines connect the characteristic temperatures from 2 temperature fits.
For clarity, the characteristic temperatures from 3 temperature fits are connected by solid lines.
Square boxes mark data dominated by flare plasma. 
Color coding as in Fig. \ref{figLxMT}.  \label{figVEM_T}}
\end{figure}

Spectral fitting of the CCD resolution X-ray spectra using XSPEC provides 
valuable information on the coronal thermal structure but is limited by the 
available X-ray signal with lower count levels providing less information.
In Fig. \ref{spectralfits}, spectra for low activity and high activity Chandra 
observations of GJ 876 are shown as examples.  
The amount of coronal plasma (volume emission measure -- VEM) as a function 
of coronal temperature measured from the CCD-resolution spectra is shown 
in Fig.  \ref{figVEM_T}. The coronal temperatures show the expected behavior 
where quiescent plasma is at cooler temperatures than flaring plasma.
The quiescent emission for all the stars shows a 
dominant VEM component peaking at a fairly cool temperature in the range 
 0.2-0.5 keV (2 - 6 MK), with many stars clustering near 0.25 keV (3 MK). 
 Note that for GJ 1132 and TRAPPIST-1 the coronal temperatures have been 
 assumed rather than measured. Flaring time intervals show higher 
 characteristic temperatures in the range 0.6-1.0 keV (7 - 12 MK).
 It should always be remembered that ``quiescent'' conditions may 
 contain low level, unrecognized flaring.

\subsection{Activity Trends of Sample}

{\it Mass and Effective Temperature:}  The plots of X-ray luminosity against mass and 
effective temperature (Fig. \ref{figLxMT}) convey essentially the same information because the two 
parameters are directly connected. As expected younger, more rapidly rotating stars show 
higher X-ray luminosities \citep{pallavicini81,walter82,pizzolato03}. 
At any particular mass or effective temperature there is a 
significant range in X-ray luminosity and some of that range can definitely be associated 
with variable flaring X-ray emission. X-rays provide a larger contribution to the overall 
bolometric luminosity for lower mass stars; a feature that becomes prominent when 
viewing the ratio of X-ray to bolometric luminosity.

{\it Rotation and Age:}
It is well known that rapid rotation drives dynamos that create the magnetic activity 
resulting in X-ray and UV radiation. This correlation has been developed using 
ground-based optical data  \citep{noyes84,west15,houdebine17} and explicitly for coronal emission 
from space-based X-ray data \citep{pizzolato03,magaudda20, johnstone21}. 
The correlation between rotation and magnetic activity extends across the 
divide between mid-M dwarfs with a radiative core and fully convective late-M dwarfs 
without any dramatic changes \citep{wright18,linsky20}. 
Initially, young M dwarfs are fast rotators showing saturated coronal emission with 
log L$_X$/L$_{bol}$ $\simeq$ -3 . Stars such as GJ 729 and GJ 674 in our sample are 
typical examples of such very active coronae. As stars grow older their rotation usually slows 
and magnetic activity declines. However, age is not a controlling factor, because an 
old but rapidly rotating star can still be magnetically active, e.g. TRAPPIST-1, 
which shows a high log L$_X$/L$_{bol}$ ratio despite being at least as old as 
the Sun, based on the kinematic and elemental abundance age estimate of  \citet{burgasser17}.

\begin{figure*}
\includegraphics[angle=90,scale=.75]{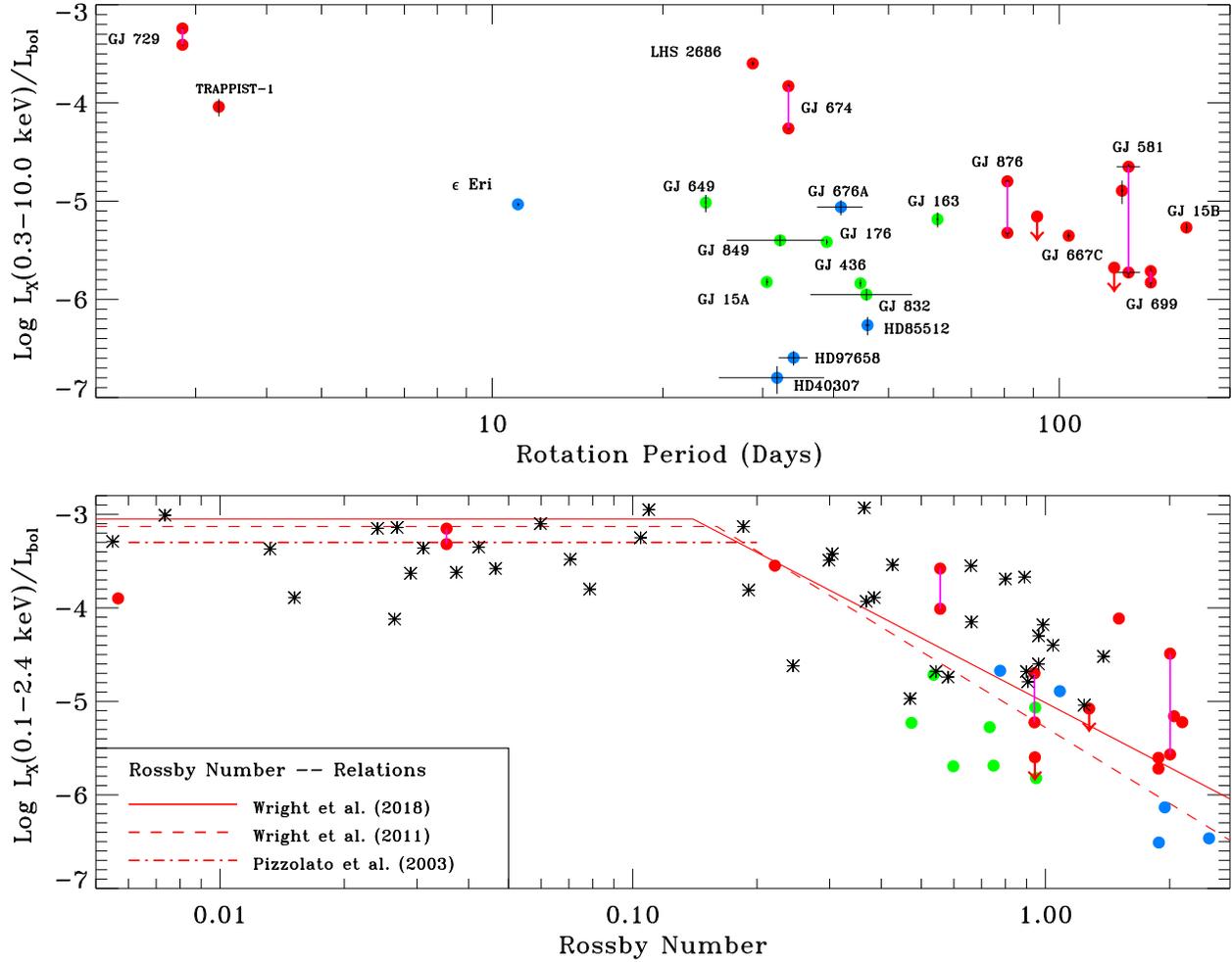}
\caption{Comparison of X-ray-to-bolometric luminosity ratio as a function of rotational period (upper panel) 
and Rossby number (lower panel).
The empirical relationships 
between the X-ray to bolometric luminosity ratio and Rossby number (rotation period/convective 
turnover timescale) derived by \citet{wright18} and \citet{wright11} from large samples of M dwarfs 
are plotted as solid and dashed lines, respectively. The saturated activity level for M dwarfs 
from \citet{pizzolato03} is shown as a dot-dash line. 
The nearby M dwarf sample from \citet{wright18} and \citet{wright11}  are shown as black asterisks. 
Color coding as in Fig. \ref{figLxMT}.  
\label{figLxRNum}}
\end{figure*}

{\it Rossby Number:}
Fig. \ref{figLxRNum} displays log-log plots comparing the ratio of the 0.3-10.0 keV 
X-ray luminosity to bolometric luminosity as a 
function of rotation period and the X-ray luminosity to bolometric luminosity in the 
ROSAT 0.1-2.4 keV energy range as a function of Rossby number. 

\citet{wright18} and \citet{wright11} performed detailed studies of the relationship 
between X-ray luminosity in the ROSAT 0.1-2.4 keV energy range and stellar 
rotation for a wide range of FGKM dwarf stars.
They found that the tightest correlations involved the Rossby number 
(R$_0$ = P$_{rot}$/$\tau$, where $\tau$ is the convective turnover time),
rather than the rotation period (P$_{rot}$) alone. The relationships discussed by 
\citet{wright18} and \citet{wright11} are plotted in Fig \ref{figLxRNum}. 
Declines from saturated emission start at R$_0$ = 0.14 and 0.16.
As a guide to the range of coronal activity detected, the sample of nearby M dwarfs included in 
those papers are also plotted. The Rossby numbers for 
all these stars were recalculated using Eqn. 5 of \citet{wright18}.
For the Rossby number plot the X-ray luminosities of our target stars were converted to the 
0.1-2.4 keV ROSAT energy range for compatibility with the other data and relationships 
shown. This energy conversion results in increased luminosities that are temperature dependent; 
at 0.5 keV the increase is 31\%, at 0.3 keV it becomes 50\%, and by 0.15 keV it is 94\%.

The MUSCLES and Mega-MUSCLES stars are consistent with previous observations of M dwarf 
coronal emission but extend the sampling to lower X-ray luminosities, larger Rossby numbers, 
and smaller masses.

\subsection{Coronal Variability}

Perhaps the most important result from the MUSCLES and Mega-MUSCLES surveys is the 
discovery that strong UV/X-ray flare activity is present on M dwarfs even down
to the oldest, slowly rotating and lowest mass stars sampled.  
Many of the stars observed, particularly the mid-M and late-M stars, show 
X-ray variability, often in the form of 
large flare outbursts and at other times as clear changes in quiescent flux 
between observations at different times. Several examples are described by 
\cite{Youngblood_etal17}, including the 
multiple flares seen by Chandra from the slowly-rotating (P$_{rot}$ = 97 days) 
M4 dwarf GJ876 (see Fig. \ref{figGJ876_GJ581}). These flares when viewed from the 
habitable zone are remarkable in comparison to present day solar flares that 
impact the Earth.

FUV and X-ray 
flares were observed that reached peaks at 10-100 times the quiescent emission on 
timescales of 10$^2$-10$^3$ seconds.  These observations were enabled by the 
temporal variability studies possible using the Cosmic Origins Spectrograph 
(COS) on HST \citep{Green_etal12} and 
the photon counting X-ray imagers on Chandra and XMM-Newton. 
COS provides both flare-related flux and spectral line profile monitoring. 
\citet{Loyd_etal18} studied the FUV variability seen in the COS 
data for the MUSCLES sample and found that most of the nominally ``inactive'' stars 
showed significant flaring in transition region and chromospheric emission lines.  
While the FUV emission declines as M dwarfs  
age and rotate more slowly, the level of flaring emission relative to quiescent 
emission remains constant.

\section{Conclusions}\label{sec:conclude}
The MUSCLES and Mega-MUSCLES HST Treasury programs have completed a 
comprehensive survey of the high energy UV/EUV/X-ray radiation fields of a 
representative sample of nearby K-M dwarf stars hosting likely terrestrial exoplanets. 
These stars sample a broad range of mass/effective temperature, rotational rates, 
and ages and should prove useful as activity proxies for more distant exoplanetary 
systems for which detailed observational studies may not be possible.

The overall conclusions concerning the coronal X-ray emission from this sample of stars 
are:
\begin{enumerate}

\item[i)]{The vast majority of the stars (21 of 23) were detected in deep 
Chandra/XMM-Newton/Swift observations and useful upper limits measured for 
the other two. Accurate X-ray luminosities were measured to quantify the coronal 
activity.This was a result of selecting the nearest suitable exoplanet 
host stars in building the sample.}

\item[ii)]{Useful spectral information was obtained for most of the detected Chandra and 
XMM-Newton sources and 
this provides estimates of the coronal characteristic temperatures. This information 
is vital for the accurate modeling of the associated EUV radiation.}

\item[iii)]{Short-term and long-term coronal variability is common, particularly among 
the M dwarfs. Significant X-ray and UV flaring is seen, not only that expected from 
young fast-rotating stars, but also from slowly rotating stars older than the Sun.}

\end{enumerate}

\begin{acknowledgments}
This work was supported by {\it Chandra} grants GO4-15014X,  
GO5-16155X, and GO8-19017X,  and NASA {\it XMM-Newton} grant NNX16AC09G 
to the University of Colorado. This work was supported by HST grants 
to Treasury programs 13650 and 15071. This research has made use of data obtained 
from the Chandra Data Archive, and software provided by the Chandra X-ray Center (
CXC) in the CIAO application package. This research has made use of data from {\it XMM-Newton}, 
an ESA science mission with instruments and contributions directly funded by ESA 
member states and NASA.
This research has made use of data and/or software provided by the High Energy 
Astrophysics Science Archive Research Center (HEASARC), which is a service of 
the Astrophysics Science Division at NASA/GSFC.
This work has made use of data from the European Space Agency (ESA) mission
{\it Gaia} (\url{https://www.cosmos.esa.int/gaia}), processed by the {\it Gaia}
Data Processing and Analysis Consortium (DPAC,
\url{https://www.cosmos.esa.int/web/gaia/dpac/consortium}). Funding for the DPAC
has been provided by national institutions, in particular the institutions
participating in the {\it Gaia} Multilateral Agreement.
\end{acknowledgments}

\vspace{5mm}
\facilities{CXO, XMM-Newton, Swift, HST(COS)}

\software{CIAO \citep{CIAO06}, XSPEC \citep{arnaud96}, IDL (Ver. 8.8; Excelis Visual Information Solutions, Boulder, CO)}


\appendix

\section{Light Curves of Non-variable Sources}

\begin{figure}[h]
\includegraphics[angle=90,scale=.70]{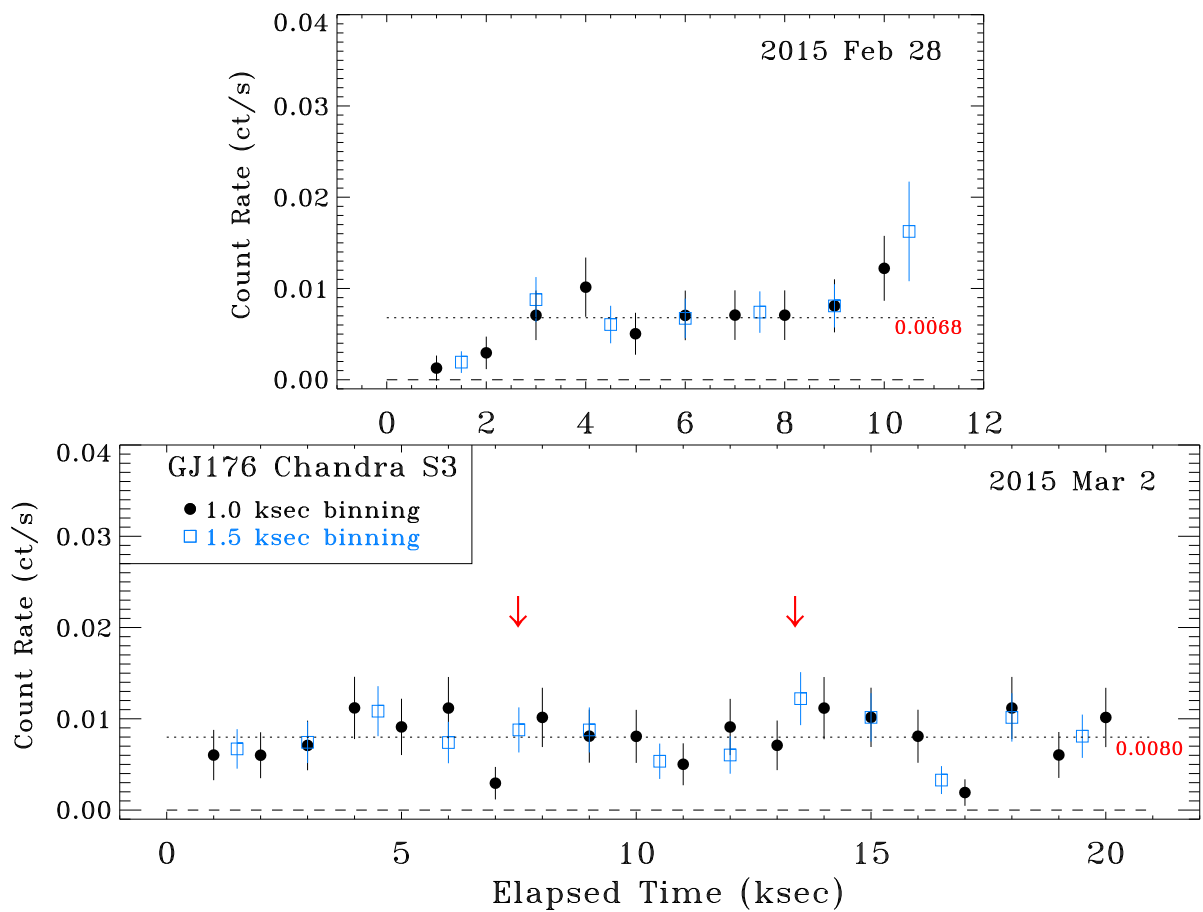}
\caption{Chandra ACIS-S3 light curves using 1.0 and 1.5 kilosecond sampling
bins from both observations of GJ 176 (M2 V). The mean count rate levels are 
shown as dotted lines and labeled in red. Dashed lines mark the zero count rate level.
Red downward arrows marks the times when  the largest FUV flares seen by  
\citet{Loyd_etal18} were detected.  1$\sigma$ error bars are plotted.   \label{figGJ176}}
\end{figure}

\begin{figure}[h]
\includegraphics[angle=90,scale=.70]{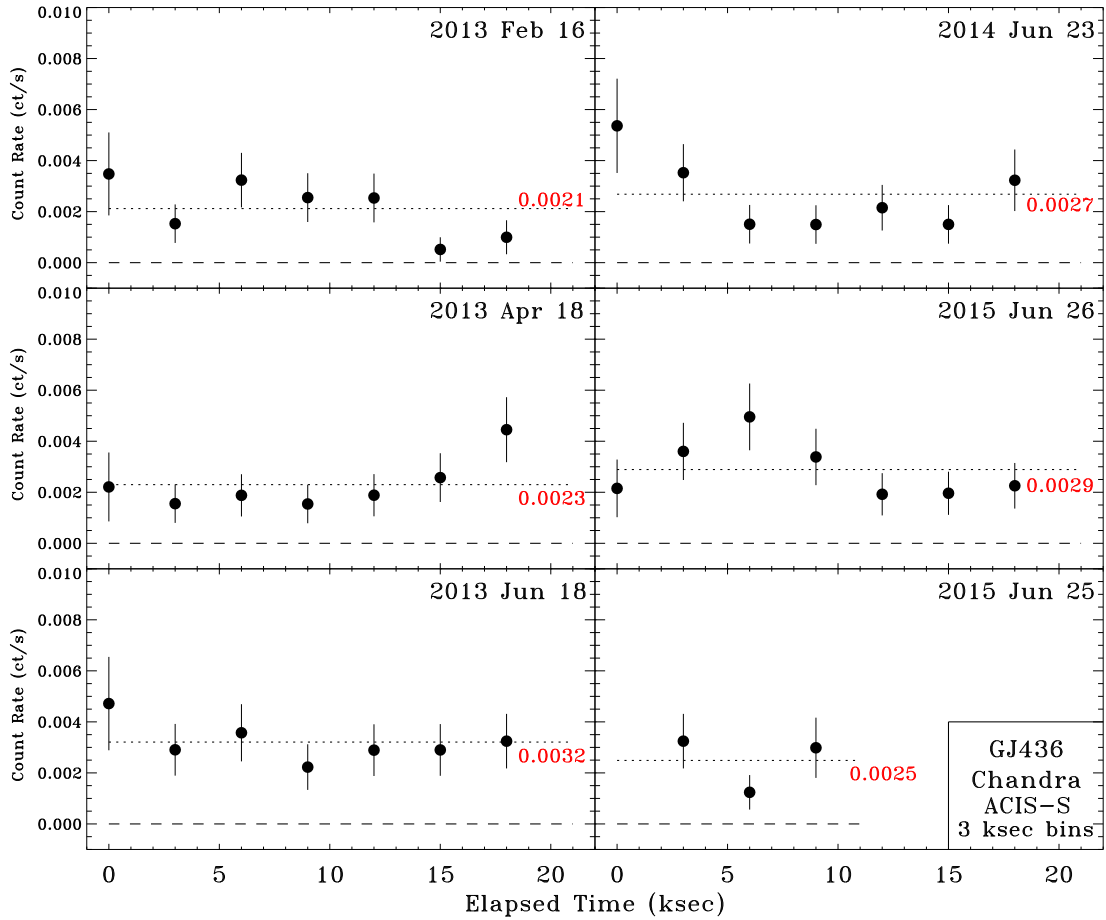}
\caption{Chandra ACIS-S3 light curves using 3 kilosecond sampling bins
from six observations of GJ 436 (M3 V). No statistical significant
variability is detected over the complete dataset, which includes a total of 
105 kiloseconds of X-ray monitoring.The mean count rates for the individual
observations are listed in the relevant panel and marked by a dotted line. 
The average of the 6 measured count rates is 0.0026$\pm$0.0004 ct/s. 
1$\sigma$ error bars are plotted. 
 \label{figGJ436}}
\end{figure}

\begin{figure}[h]
\includegraphics[angle=90,scale=.53]{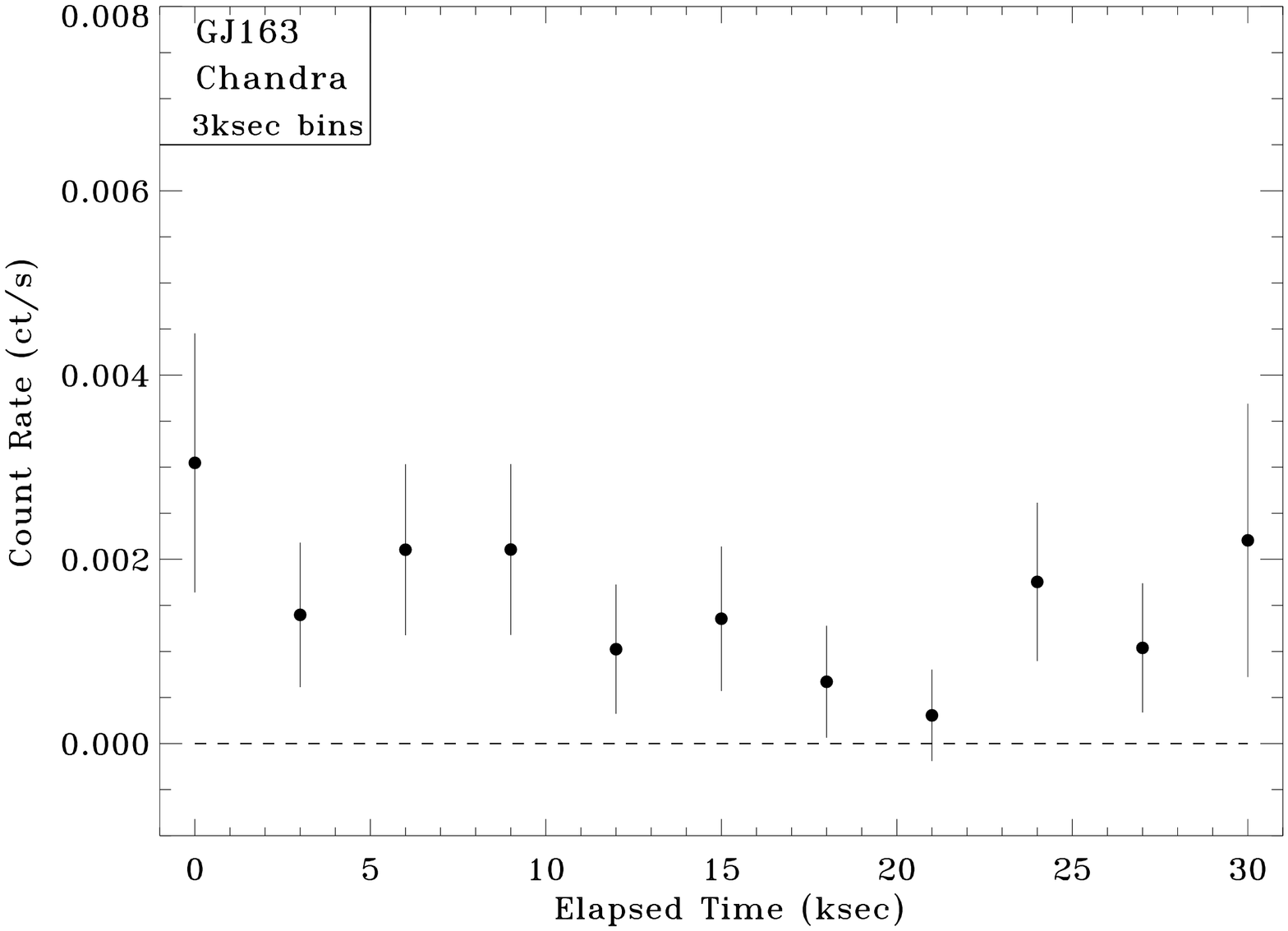}
\caption{Chandra ACIS-S3 light curve of GJ 163 (M3.5 V, 0.405 M$_\odot$) 
using 3 kilosecond sampling bins. 1$\sigma$ error bars are plotted. 
No statistically significant variability is detected with 
 a glvary VARINDEX = 2.  \label{figGJ163}}
\end{figure}

\begin{figure}[h]
\includegraphics[angle=90,scale=.53]{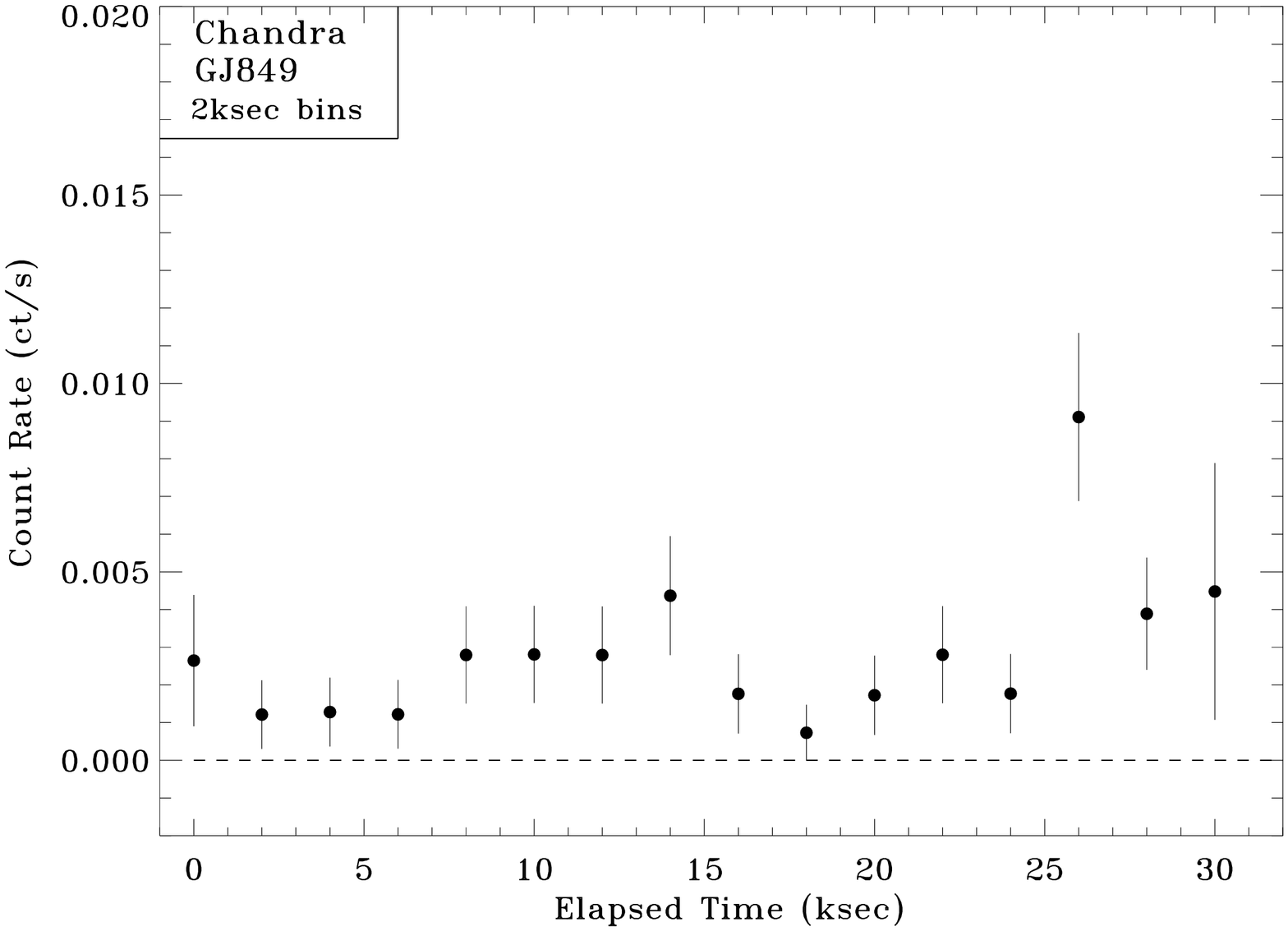}
\caption{Chandra ACIS-S3 light curve of GJ 849 (M3.5 V, 0.465 M$_\odot$) 
using 2 kilosecond sampling bins. 1$\sigma$ error bars are plotted. 
No statistically significant variability is detected with 
 a glvary VARINDEX = 0. \label{figGJ849}}
\end{figure}


\clearpage

\begin{thebibliography}{}
\bibitem[Afram \& Berdyugina(2019)]{Afram_Berdyuginal19}
Afram, N. \& Berdyugina, S.V. 2019, \aap, 629, A83
 \bibitem[Allen \& Herrera(1998)]{allen98} Allen, C. \& Herrera, M. A. 1998, \rmxaa, 34, 37
 \bibitem[Alonso-Floriano et al.(2015)]{alonso-floriano15}
 Alonso-Floriano, F. J., Morales, J. C., Caballero, J. A.,  et al. 2015, \aap, 577, A128
 \bibitem[Anglada-Escud\'{e} et al.(2013)]{anglada-escude13}Anglada-Escud\'{e}, G., Tuomi, M., 
 Gerlach, E.,  et al. 2013,  \aap, 556, A126

\bibitem[Arnaud (1996)]{arnaud96} Arnaud, K. A. 1996, {\it in} Astronomical
Data Analysis Software and Systems V, eds. G. Jacoby and J. Barnes, ASP Conf.
Series Vol. 101, p.17
\bibitem[Berta-Thompson et al.(2015)]{berta-thompson15}Berta-Thompson, Z. K., Irwin, J., Charbonneau, D., 
et al 2015, \nat, 527, 2004.
\bibitem[Bonfanti et al.(2015)]{bonfanti15} Bonfanti, A., Ortolani, S., Piotto, G., \& 
Nascimbeni, V. 2015, \aap, 575, A18
\bibitem[Bonfils et al.(2013)]{bonfils13}Bonfils, X., Lo Curto, G., Correia, A. C. M., et al 
2013, \aap, 556, A110
\bibitem[Bryden et al.(2009)]{bryden09}Bryden, G., Beichman, C. A., Carpenter, J. M., et al. 2009,
\apj, 705, 1226
\bibitem[Burgasser \& Mamajek(2017)]{burgasser17}
Burgasser, A. J. \& Mamajek, E. E. 2017, \apj, 845, 110
\bibitem[Chabrier \& Baraffe(1997)]{chabrier97}Chabrier, G.  \& Baraffe, I. 1997, \aap, 327, 1039
\bibitem[Coffaro et al.(2020)]{coffaro20} Coffaro, M., Stelzer, B., Orlando, S., et al. 2020, \aap, 636, A49
\bibitem[D\'{i}ez Alonso et al.(2019)]{diez_alonso19}D\'{i}ez Alonso, E., Caballero, J. A., Montes, D. et al. 
2019, \aap, 621, A126
\bibitem[Donahue, Saar \& Baliunas(1996)]{donahue96} Donahue, R. A., Saar, S. H., 
\& Baliunas, S. L.  1996, \apj, 466, 384
\bibitem[Dorman, Arnaud, \& Gordon(2003)]{dorman03} Dorman, B., Arnaud,
K. A., \& Gordon, C. A. 2003, \baas 35, 641
\bibitem[Dressing \& Charbonneau(2015)]{dressing15}Dressing, C. D. \& Charbonneau, D., 
2015, \apj, 807, 45
\bibitem[Duvvuri et al.(2021)]{duvvuri21}Duvvuri, G. M., Pineda, J. S., Berta-Thompson, Z. K., 
et al. 2021, \apj, 913, 40
\bibitem[Fleming et al.(1995)]{fleming95}Fleming, T. A., Schmitt, J. H. M. M., \& Giampapa, M. S. 
1995, \apj, 450, 401
\bibitem[France et al.(2020)]{france20}France, K., Duvvuri, G., Egan, H., et al. 2020, \aj, 160, 237
\bibitem[France et al.(2013)]{france13}France, K., Froning, C. S., Linsky, J. L., et al. 2013, \apj, 763, 149
\bibitem[France et al.(2012)]{france12}France, K., Linsky, J. L., Tian, F., et al. 2012, \apjl, 750, L32
\bibitem[France et al.(2016)]{france16}France, K., Loyd, R. O. P., Youngblood, A., et al. 2016, \apj, 820, 89
\bibitem[Froning et al.(2019)]{froning19}Froning, C. S., Kowalski, A., France, K. et al. 
2019, \apjl, 871, L26
\bibitem[Froning et al.(2022)]{froning22}Froning, C. S., Wilson, D., France, K. et al. 
2022, \baas, 54e.023F
\bibitem[Fruscione et al.(2006)]{CIAO06}Fruscione, A., McDowell, J. C., Allen, G. E., et al. 2006,
SPIE, Vol. 6270, id. 62701V
\bibitem[Gaia Collaboration et al.(2016)]{GAIA16}  Gaia Collaboration et al. 2016,  \aap, 595,A1
\bibitem[Gaia Collaboration et al.(2020)]{GAIA20} Gaia Collaboration et al.  2020,   arXiv:2012.01533
\bibitem[Gonzales et al.(2019)]{Gonzales_etal19}
Gonzales, E. C., Faherty, J. K., Gagn\'{e}, J. , et al.  2019, \apj, 886, 131
\bibitem[Green et al.(2012)]{Green_etal12}
Green, J.C., Froning, C.S., Osterman, S.,  et al.  2012, \apj, 744, 60
\bibitem[G\"{u}del(2007)]{gudel07} G\"{u}del, M. 2007, LRSP., 4, 3
\bibitem[Guo et al.(2020)]{guo20}Guo, X., Crossfield, I. J. M., Dragomir, D., et al. 
2012, \aj, 159, 239
\bibitem[Hawley, Gizis, \& Reid(1996)]{hawley96} Hawley, S. L., Gizis, J. E., \& 
reid, I. N. 1996, \aj, 112, 2799
\bibitem[ Houdebine et al.(2017)]{houdebine17}Houdebine, E. R., Mullan, D. J., Bercu, B., et al. 
2017, \apj, 837, 56
\bibitem[ Houdebine et al.(2019)]{houdebine19}Houdebine, E. R., Mullan, D. J., Doyle, J. G. et al. 
2019, \aj, 158, 96
\bibitem[Iba\~{n}ez Bustos et al.(2020)]{Ibanez_Bustos20} Iba\~{n}ez Bustos, R. V., Buccino, A. P., Messina, S., 
Lanza, A. F., \& Mauas, P. J. D. 2020, \aap, 644, A2
\bibitem[Janson et al.(2008)]{janson08} Janson, M., Reffert, S.,
Brandner, W., et al., 2008,  \aap, 488, 771
\bibitem[Johnstone, Bartel, \& G\"{u}del(2021)]{johnstone21}Johnstone, C. P., Bartel, M., \& G\'{u}del, M. 
2021, \aap, 649, A96
\bibitem[Kiraga \& Stepien(2007)]{kiraga07} Kiraga, M. \& Stepie\'{n}, K. 2007, \actaa, 57, 149
\bibitem[Koen et al.(2010)]{Koen_etal10}Koen, C., Kilkenny, D., van Wyk, F., \& Marang, F. 
2010, \mnras, 403,1949
\bibitem[Kulenthirarajah et al.(2017)]{kulen17}
Kulenthirarajah, L., Donati, J.-F., Hussain, G., Morin, J., \& Allard, F. 2017, \mnras, 487, 1335
\bibitem[Linsky et al.(2020)]{linsky20}Linsky, J. L., Wood, B. E., Youngblood, A., et al.
2020, \apj, 902,3

\bibitem[Loyd et al.(2016)]{Loyd_etal16}
Loyd, R. O. P., France, K., Youngblood, A., et al.  2016, \apj, 824, 102
\bibitem[Loyd et al.(2018)]{Loyd_etal18}
Loyd, R. O. P., France, K., Youngblood, A., et al.  2018, \apj, 867, 71
\bibitem[Magaudda et al.(2020)]{magaudda20}Magaudda, E., Stelzer, B., Covey, K. R., et al. 
2020, \aap, 638, A20
\bibitem[Malkov et al.(2012)]{malkov12}Malkov, O. Yu., Tamazian, V. S., Docobo, J. A., \& 
Chulkov, D. A. 2012, \aap, 546, 69
\bibitem[Mallonn et al.(2018)]{mallonn18}
Mallonn, M., Herrero, E., Juvan, I. G., et al.  2018, \aap, 614, A35
\bibitem[Mann et al.(2019)]{mann19}Mann, A. W., Dupuy, T., Kraus, A. L., et al. 2019, \apj, 871, 63
\bibitem[Metcalfe et al.(2013)]{metcalfe13}Metcalfe, T. S., Buccino, A. P., Brown, B. P., et al. 
2013, \apjl, 763, L26
\bibitem[Montes et al.(2001)]{montes01} Montes, D., L\'{o}pez-Santiago, J., G\'{a}lvez, M. C., et al. 
2001, \mnras, 328, 45
\bibitem[Newton et al.(2014)]{newton14}Newton, E. R., Charbonneau, D., Irwion, J., et al.,
2014, \aj, 147, 20
\bibitem[Newton et al.(2016)]{newton16}Newton, E. R., Irwin, J., Charbonneau, D., et al.,
2016, \apj, 821, 93
\bibitem[Newton et al.(2018)]{newton18}Newton, E. R., Mondrik, N., Irwin, J., Winters, J. G., \&
Charbonneau, D. 2018, \aj, 156, 217
\bibitem[Noyes et al.(1984)]{noyes84}Noyes, R. W., Hartmann, L. W., Baliunas, S. L., 
Duncan, D. K., \& Vaughan, A. H. 1984, \apj, 279, 763
\bibitem[Pallavicini et al.(1981)]{pallavicini81}Pallavicini, R., Golub, L., Rosner, R., et al. 
1981, \apj, 248, 279
\bibitem[Pinamonti et al.(2018)]{pinamonti18}Pinamonti, M., Damasso, M., Marzari, F., et al. 
2018, \aap, 617, A104 
\bibitem[Pineda, Youngblood, \& France(2021)]{pineda21b}Pineda, J. S., Youngblood, A. \& 
France, K. 2021b, \apj, 918, 40
\bibitem[Pizzolato et al.(2003)]{pizzolato03}Pizzolato, N., Maggio, A., Micela, G., Sciortino, S., \& 
Ventura, P. 2003, \aap, 397, 147
 \bibitem[Poppenhaeger, Robrade, \& Schmitt(2010)]{poppenhaeger10} 
 Poppenhaeger, K., Robrade, J., \& Schmitt, J. H. M. M. 2010, \aap, 515, A98
\bibitem[Reid, Hawley, \& Gizis(1995)]{reid95} Reid, I. N., Hawley, S. L., \& 
Gizis, J. E. 1995, \aj, 110, 1838
\bibitem[Ribas et al.(2005)]{ribas05} Ribas, I., Guinan, E. F., G\"{u}del, M., \& Audard, M.  2005, 
\apj, 622, 680
\bibitem[Ribas et al.(2018)]{ribas18} Ribas, I.,  Tuomi, M., Reiners, A. et al. 2018, 
\nat, 563, 365
\bibitem[Rivera et al.(2005)]{rivera05} Rivera, E. J., Lissauer, J. J., Butler, R. P., et al.
2005, \apj, 634, 625
\bibitem[Robertson, Mahadevan, Endl, \& Roy(2014)]{robertson14a}
Robertson, P., Mahadevan, S., Endl, M., \& Roy, A. 2014, Science, 345, 440
\bibitem[Schweitzer et al.(2019)]{Schweitzer_etal19}
Schweitzer, A., Passegger, V. M., Cifuentes, C. et al. 2019, \aap, 625, A68
\bibitem[Shulyak et al.(2019)]{Shulyak_etal19}
Shulyak, D., Reiners, A., Nagel, E., et al.   2019, \aap, 626, A86
\bibitem[Smith et al.(2001)]{smith01}Smith, R. K., Brickhouse, N. S., Liedahl, D. A., \& Raymond, J. C. 2001, \apjl, 556, L91
\bibitem[Sousa et al.(2008)]{sousa08} Sousa, S. G., Santos, N. C., Mayor, M., et al. 
2008, \aap, 487, 373
\bibitem[Su\'{a}rez Mascare\~{n}o et al.(2015)]{mascareno15}
Su\'{a}rez Mascare\~{n}o, A., Rebolo, R., Gonz\'{a}lez Hern\'{a}ndez, J. I., 
Esposito, M.  2015, \mnras, 452, 2745
 \bibitem[Takeda \& Honda(2020)]{takeda20} Takeda , Y. \& Honda, S. 
 2020, \aj, 159, 174
\bibitem[Toledo-Padr\'{o}n et al. (2019)]{toledo-padron19} Toledo-Padr\'{o}n, B., 
Gonz\'{a}lez hern\'{a}ndez, J. I., Rodr\'{i}guez-L\'{o}pez, et al. 2019, \mnras, 488, 5145
\bibitem[Tuomi et al.(2013)]{tuomi13}
Tuomi, M., Anglada-Escud\'{e}, G., Gerlach, E., et al. 2013, \aap, 549, A48
\bibitem[Van Grootel et al.(2014)]{vanGrootel14} Van Grootel, V., Gillon, M., 
Valencia, D., et al. 2014, \apj, 786, 2
\bibitem[Veyette \& Muirhead(2019)]{veyette18}Veyette, M. J. \& Muirhead, P. S. 
2019, \apj, 863, 166
\bibitem[Vida et al.(2017)]{Vida17}Vida, K., K\"{o}v\'{a}ri, Zs., P\'{a}l, A., Ol\'{a}h, \& 
Kriskovics, L.  2017, \apj, 841, 124
\bibitem[Walter(1982)]{walter82}Walter, F. M. 1982, \apj, 253, 745
\bibitem[Wargelin et al.(2008)]{wargelin08}Wargelin, B. J., Kashyap, V. L., 
Drake, J. J., Garc\'{i}a-Alvarez, D., \& Ratzlaff, P. W.  2008, \apj, 676, 610
\bibitem[West et al.(2015)]{west15}West, A. A., Weisenburger, K. L., Irwin, J., 
at al. 2015, \apj, 812, 3
\bibitem[Wheatley et al.(2017)]{wheatley17}Wheatley, P. J., Louden, T., 
Bourrier, V., Ehrenreich, D., \& Gillon, M. 2017, /mnras, 465, L74
\bibitem[Wilson et al.(2021)]{wilson21}Wilson, D. \& The Mega-MUSCLES Collaboration 2021, 
https://doi.org/10.5281/zenodo.4567579
\bibitem[Wright et al.(2011)]{wright11}Wright, N. J., Drake, J. J., Mamajek, E. E., 
 \& Henry, G. W.  2011, \apj, 743, 48
\bibitem[Wright et al.(2018)]{wright18}Wright, N. J., Newton, E. R., Williams, P. K. G.
Drake, J. J., \& Yadav, R. K. 2018, \mnras, 479, 2351
\bibitem[Youngblood et al.(2017)]{Youngblood_etal17}
Youngblood, A., France, K., Loyd, R.O.P., et al.  2017,  \apj, 843, 31
\bibitem[Youngblood et al.(2016)]{Youngblood_etal16}
Youngblood, A., France, K., Loyd, R.O.P., et al.  2016, \apj, 824, 101

\end{thebibliography}



\end{document}